\documentclass[12pt]{article}
\usepackage{amsmath}
\usepackage{graphicx,psfrag,epsf}
\usepackage{enumerate}
\usepackage{natbib}
\usepackage{multirow}
\usepackage{setspace}
\usepackage{color}
\usepackage{bm}
\usepackage{subfig}
\usepackage{caption}

\newcommand{\blind}{0}

\begin{document}


\if0\blind
{
  \title{\bf A comparison of efficient permutation tests for unbalanced ANOVA in two by two designs--and their behavior under heteroscedasticity}
  \author{Sonja Hahn
    \\Department of Psychology, University of Jena\\
    and \\
    Frank Konietschke\\
    Department of Medical Statistics, University Medical Center G\"{o}ttingen\\
    and\\
    Luigi Salmaso \\
    Department of Management and Engineering, University of Padova}
  \maketitle
} \fi

\if1\blind
{
  \bigskip
  \bigskip
  \bigskip
  \begin{center}
    {\LARGE\bf Title}
\end{center}
  \medskip
} \fi

\bigskip
\begin{abstract}
We compare different permutation tests and some parametric counterparts that are applicable to unbalanced designs in two by two designs. First the different approaches are shortly summarized. Then we investigate the behavior of the tests in a simulation study. A special focus is on the behavior of the tests under heteroscedastic variances.
\end{abstract}

\noindent%
\vfill

\newpage
\doublespacing

\section{Introduction}
\label{sec:intro}

In many biological, medical and social trials, data are collected in terms of a two by two design, e.g. when male and female patients are randomized to two different treatment groups (placebo and active treatment). The data is often analyzed by assuming linear treatment effects and ANOVA procedures.  These approaches rely on rather strict model assumptions like normally distributed error terms and variance homogeneity. However, these model assumptions can rarely be justified. In particular, heteroscedastic variances occur frequently in a variety of disciplines, e.g. in genetic data. It is well known that the classical ANOVA $F$-test tends to result in liberal or conservative decisions, depending on the underlying distribution, the amount of variance heterogeneity, and unbalance. Thus, asymptotic (or approximate) procedures, which allow the data to be heteroscedastic, are a robust alternative to the classical ANOVA $F$-test. An asymptotic testing procedure is the Wald-type statistic \citep[see e.g.,][]{di250,di242}, which is based on the asymptotic distribution of an appropriate quadratic form. It is even valid without the assumptions of normality and variance homogeneity. However, very large sample sizes are necessary to achieve accurate test results (see, e.g., \cite{di250} and references therein). As an approximate solution, \cite{di250} propose the so-called ANOVA-type statistic (ATS), which is based on an Box-type approximation approach. The ATS, however, is an approximate test and its asymptotically exactness is unknown (see, e.g., \cite{di242}). On the other hand, permutation approaches are known to be very robust under non-normality. In particular, under certain model assumptions, permutation tests are exact level $\alpha$ tests. Usual permutation tests assume that the data is exchangeable, which particularly implies homogeneous variances. Recently, Pauly et al. (2013) propose asymptotic permutation tests, which are asymptotically exact even under non-normality and possibly heteroscedastic variances. \\
Various permutational approaches for factorial designs have been developed within the last years, but a comparison of the different permutational approaches for unbalanced factorial designs with variance heterogeneity remains.\\
The aim of the present paper is to investigate different parametric and permutation tests for factorial linear models. For simplicity, we focus on two by two designs within this paper. \\
The paper is organized as follows: After some notational issues we summarize different existing approaches that were developed for unbalanced ANOVA designs. Afterwards we investigate the behavior of these procedures in a simulation study. Here we focus on small sample sizes, heterogeneity of variances, and different error term distributions. Finally, we discuss the results of the simulation study and add further considerations about the procedures.

\subsection{Notation and Hypotheses}
\label{subsec:notation}
We consider the two way factorial crossed design
\begin{eqnarray} \label{model}
X_{ijk}= \mu + \alpha_i +\beta_j +(\alpha\beta)_{ij} +\epsilon_{ijk},\; i=1,2;\; j=1,2;\; k=1,\ldots,n_{ij},
\end{eqnarray}
where $\alpha_i$ denotes the effect of level $i$ from factor A, $\beta_j$ denotes the effect of level $j$ from factor $B$ and $(\alpha\beta)_{ij}$ denotes the $(ij)$th interaction effect from $A\times B$. Here, $\epsilon_{ijk}$ denotes the error term with $E(\epsilon_{ijk})=0$ and $Var(\epsilon_{ijk})=\sigma_{ij}^2 >0$. Under the assumption of equal variances, we simply write $Var(\epsilon_{ijk})=\sigma^2$. It is our purpose to test the null hypotheses
\begin{eqnarray} \label{hypotheses}
H_0^{(A)}&:&\alpha_1=\alpha_2 \label{H_0_A}\nonumber\\
H_0^{(B)}&:&\beta_1=\beta_2 \label{H_0_B}\\
H_0^{(A\times B)}&:&(\alpha \beta)_{11}=\ldots=(\alpha \beta)_{22} \label{H_0_AB} \nonumber
\end{eqnarray}
For simplicity, let $\mu_{ij}=\mu+\alpha_i+\beta_j+(\alpha\beta)_{ij}$, then, the hypotheses defined above can be equivalently  written as
\begin{eqnarray*}
H_0^{(A)}&:&\mathbf{C}_A\bm{\mu}=\mathbf{0}\\
H_0^{(B)}&:&\mathbf{C}_B\bm{\mu}=\mathbf{0} \\
H_0^{(A\times B)}&:&\mathbf{C}_{A\times B}\bm{\mu}=\mathbf{0},
\end{eqnarray*}
where $\mathbf{C}_L, L\in \{A,B,A \times B\}$, denote suitable contrast matrices and $\bm{\mu}=(\mu_{11},\ldots,\mu_{22})'$.
To test the null hypotheses formulated in (\ref{hypotheses}), various asymptotic and approximate test procedures have been proposed. We will explain the current state of the art in the subsequent sections.

\subsection{Wald-Type Statistic (WTS)}
\label{subsec:waldparam}

Let $\mathbf{\overline{X}}_\cdot = (\overline{X}_{11\cdot},\ldots,\overline{X}_{22\cdot})'$ denote the vector of sample means $\overline{X}_{ij\cdot}=\frac{1}{n_{ij}}\sum_{k=1}^{n_{ij}} X_{ijk}$, and let $\widehat{\mathbf{S}}_N = diag(\widehat{\sigma}_{11}^2,\ldots,\widehat{\sigma}_{22}^2)$ denote the $4 \times 4$ diagonal matrix of sample variances $\widehat{\sigma}_{ij}^2= \frac{1}{n_{ij}-1}\sum_{k=1}^{n_{ij}}(X_{ijk} - \overline{X}_{ij\cdot})^2$.
Under the null hypothesis $H_0:(L):\mathbf{C}_L\bm{\mu}=\mathbf{0}$, the Wald-type statistic
\begin{eqnarray} \label{wald}
W_N(L)=N\mathbf{\overline{X}}_\cdot'\mathbf{C}_L'(\mathbf{C}_L \mathbf{S}_N \mathbf{C}_L')^+ \mathbf{C}_L\mathbf{\overline{X}}_\cdot \to \chi_{rank(\mathbf{C}_L)}^2
\end{eqnarray}
has, asymptotically, as $N\to \infty$, a $\chi_{rank(\mathbf{C}_L)}^2$ distribution. The rate of convergence, however, is rather slow, particularly for larger numbers of factor levels and smaller sample sizes. For small and medium sample sizes, the WTS tends to result in rather liberal results \citep[see][for some simulation results]{di250,di242}. However, the Wald-type statistic is asymptotically exact even under non-normality.  

\subsection{ANOVA-Type Statistic (ATS)}
\label{subsec:ATSparam}
In order to overcome the strong liberality of the Wald-type statistic in (\ref{wald}) with small sample sizes, \cite{di250} propose the so-called ANOVA-type statistic (ATS)

\begin{eqnarray}
F_N(L) &=& \frac{N \ \mathbf{\overline{X}}_\cdot' \mathbf{T}_L \mathbf{\overline{X}}_\cdot}{trace(\mathbf{T}_L \mathbf{S}_N)}, \label{FNT}
\end{eqnarray}
where $\mathbf{T}_L=\mathbf{C}_L'(\mathbf{C}_L\mathbf{C}_L')^+\mathbf{C}_L$.
The null distribution of $F_N(L)$ is approximated by a $F$-distribution with
\begin{eqnarray}
 f_1 \ = \ \frac{[trace(\mathbf{T}_L\mathbf{S}_N )]^2}{trace[(\mathbf{T}_L \mathbf{S}_N)^2]} & \text{and} &
 f_2 \ = \ \frac{[trace(\mathbf{T}_L \mathbf{S}_N)]^2}{trace(\mathbf{D}_{T_L}^2 \mathbf{S}_N^2 \mathbf{\Lambda})} \ ,
\label{atsdf}
\end{eqnarray}
where $\mathbf{D}_T$ is the diagonal matrix of the diagonal elements of $\mathbf{T}_L$ and $\mathbf{\Lambda} = diag\{ (n_{ij}-1)^{-1} \}_{i,j=1,2}$. The ATS relies on the assumption of normally distributed error terms \citep{di242}. Especially for skewed error terms the procedure tends to be very conservative \citep{di242,di253}. When the sample sizes are extremely small ($n_{ij}\approx 5$), it tends to result in conservative decisions \citep{di251}. We note that in two by two designs, the Wald-type statistics $W_N(L)$ in (\ref{wald}) and the ANOVA-type statistic $F_N(L)$ are identical. Furthermore, the ATS is even asymptotically an approximate test and its asymptotical exactness is unknown.

\subsection{Wald-Type Permutation Test (WTPS)}
\label{subsec:waldperm}
Recently, \cite{di242} proposed an asymptotic permutation based Wald-test, which is even asymptotically exact when the data is not exchangeable. In particular, it is asymptotically valid under variance heterogeneity. This procedure denotes an generalization of two-sample studentized permutation tests for the Behrens-Fisher problem \citep{Janssen1997, JanssenPauls2003, KP2012, KP2013}. The procedure is based on (randomly) permuting the data $\mathbf{X}^\ast=(X_{111}^\ast,\ldots,X_{22n_{22}}^\ast)'$ within the whole data set. Let $\overline{\mathbf{X}}_\cdot^\ast = (\overline{X}_{11\cdot}^\ast,\ldots,\overline{X}_{22\cdot}^\ast)'$ denote the vector of permuted means $\overline{X}_{ij\cdot}^\ast = n_{ij}^{-1}\sum_{k=1}^{n_{ij}}X_{ijk}^\ast$, and let $\widehat{\mathbf{S}}^\ast_N = diag(\widehat{\sigma}_{11}^{2\ast},\ldots,\widehat{\sigma}_{22}^{2\ast})$ denote the $4 \times 4$ diagonal matrix of permuted sample variances $\widehat{\sigma}_{ij}^{2\ast}= \frac{1}{n_{ij}-1}\sum_{k=1}^{n_{ij}}(X_{ijk}^\ast - \overline{X}_{ij\cdot}^\ast)^2$. Further let
\begin{eqnarray}
W_N^\ast(L)=N(\mathbf{\overline{X}}_\cdot^{\ast})'\mathbf{C}_L'(\mathbf{C}_L \mathbf{S}_N^\ast \mathbf{C}_L')^+ \mathbf{C}_L\mathbf{\overline{X}}_\cdot^\ast
\end{eqnarray}
denote the permuted Wald-type statistics $W_N(L)$. \cite{di242} show that, given the data $\mathbf{X}$, the distribution of $W_N^\ast(L)$ is, asymptotically, the $\chi_{rank(\mathbf{C}_L)}^2$ distribution. The $p$-value is derived as the proportion of test statistics of the permuted data sets that are equal or more extreme than the test statistic of the original data set. \\
If data is exchangeable, this Wald-type permutation tests guarantees an exact level $\alpha$ test. Otherwise, this procedure is asymptotically exact due to the multivariate studentization. Simulation results showed that this tests adheres better to the nominal $\alpha$-level than its unconditional counterpart for small and medium sample sizes \citep[see][and the supplementary materials therein]{di242}. Furthermore, the Wald-type permutation tests achieves a higher power than the ATS in general. We note that the WTPS is not restricted to two by two designs. The procedure is applicable in higher-way layouts and even in nested and hierarchical designs.

\subsection{Synchronized Permutation Tests (CSP and USP)}
\label{subsec:synchr}
Synchronized permutation tests were designed to test the different hypotheses in (\ref{hypotheses}) of a factorial separately (e.g., testing a main effect when there is an interaction effect). There are two important differences to the WTPS approach: (1) Data is not permuted within the whole data set, but there exists a special synchronized permutation mechanism. (2) The test statistic is not studentized. These procedure assumes that the error terms are exchangeable.\\
\citet{di109}, \citet{di039} and \citet{di112} propose synchronized permutation tests for balanced factorial designs.
\emph{Synchronization} means that data is permuted within blocks built by one of the factors. In addition,  the number of exchanged observations in each of these blocks is equal for a single permutation. For example when testing for the main effect A or the interaction effect, the observations can be permuted within the blocks built by the levels of factor B. Different variants of synchronized permutations have been developed \citep[see][for details]{di191}:
\begin{description}
  \item[Constrained Synchronized Permutations (CSP)] Here only observations on the same position within each subsample are permuted. When applied to real data set it is strongly recommended to pre-randomize the observations in the data set to eliminate possible systematic order effects.
  \item[Unconstrained Synchronized Permutations (USP)] Here also observations on different position can be permuted. In this case it has to be ensured that the test statistic follows a uniform distribution.
\end{description}
The test statistics for the main effect A and the interaction effect are
\begin{eqnarray*}
T_A&=&(T_{11}+T_{12}-T_{21}-T_{22})^2\;\;\; \text{, and}\\
T_{A\times B}&=&(T_{11}-T_{12}-T_{21}+T_{22})^2
\end{eqnarray*}
with
\begin{eqnarray*}
T_{ij}&=\sum_k X_{ijk}.
\end{eqnarray*}
Due to the synchronization and the test statistic, the effects not of interest are eliminated \citep[e.g, when testing for main effect A, main effect B and the interaction effect are eliminated, see][for more background information]{di109}.When testing for main effect B, the data has to be permuted within blocks built by the levels of A and the test statistics have to be adapted.\\
For certain unbalanced factorial designs this method can be extended \citep{di241}. In the case of CSP this leads to the situation that some observations will never be exchanged. In the case of USP the maximum number of exchanged observations equals the minimum subsample size. \\
A test statistic that finally eliminates the effects of interest is only available in special cases \citep{di241}. For example, when $n_{11}=n_{12}$ and $n_{21}=n_{22}$, possible test statistics are:
\begin{eqnarray*}
T_A&=(n_{21} T_{11}+ n_{22} T_{12}- n_{11} T_{21}- n_{12} T_{22})^2,\\
T_{A\times B}&=(n_{21} T_{11}- n_{22} T_{12}- n_{11} T_{21}+ n_{12} T_{22})^2.
\end{eqnarray*} \\
For both, the balanced and the unbalanced case, the $p$-value is again calculated as the proportion of test statistics of permuted data sets greater or equal than the test statistic of the original data set.
\\
These procedures showed a good adherence to the nominal $\alpha$-level as well as power in simulation studies \citep{di109,di241}.
However, this procedure is limited in various ways:
\begin{itemize}
   \item It is restricted to very specific cases of unbalanced designs due to assumptions on equal sized subsamples.
   \item Extension to more complex factorial designs seems quite difficult (see e.g., \citet{di109} for balanced cases with more levels).
   \item It assumes exchangeability. This might not be given in cases with heteroscedastic error variances.
\end{itemize}
As the behavior of this procedure under variance heterogeneity has not been investigated yet, we included it in the following simulation study.

\subsection{Summary}

We outlined various procedures that aim to compensate shortcomings of classical ANOVA. Some procedures are only valid under normality and possibly heteroscedastic variances (ATS). CSP and USP are valid under non normally distributed error terms and homoscedastic variances. Both the WTS and WTPS are asymptotically valid even under non-normality and heteroscedasticity, respectively. Most of these procedures are intended to be used for small samples (ATS, WTPS, CSP, and USP), only the WTS requires a sufficiently large sample size.\\
In the following simulation we vary additionally the aspect of balanced vs. unbalanced designs, as heteroscedasticity is especially problematic in the latter one.

\section{Simulation study}
\label{sec:simstudy}

\subsection{General aspects}
\label{subsec:simsdesign}

The present simulation study investigates the behavior of the procedures described above (see Section \ref{sec:intro}) for balanced vs. unbalanced designs and homo- vs. heteroscedastic variances. A major assessment criterion for the accuracy of the procedures is their behavior when increasing
sample sizes are combined with increasing variances (positive pairing) or with decreasing
variances (negative pairing).\\
We investigate data sets that did not contain any effect, and data sets that contained an effect. In the first case we were interested if the procedures keep the nominal level; in the second case additionally the power behavior was investigated. Similar to the notation introduced above we used the following approach for data simulation:
\begin{eqnarray}
X_{ijk}= \mu + \alpha_i +\beta_j +(\alpha\beta)_{ij} +\epsilon_{ijk}. \label{eq_data}
\end{eqnarray}
Specifications for the different data settings can be found below.
Throughout all studies we focused at main effect A and the interaction effect.\\
All simulations were conducted using the freely available software \textit{R} (www.r-project.org), version 2.15.2 \cite{di249}. The numbers of simulation and permutation runs were $n_{sim}=5000$ and $n_{perm}=5000$, respectively. All simulations were conducted at $5\%$ level of significance.

\subsection{Data Sets Containing no Effect}
\label{subsec:NoEffect}

\subsubsection{Description}
Table \ref{design} outlines the combinations of balanced vs. unbalanced designs and homo- vs. heteroscedastic variances. Larger sample sizes were obtained by adding a constant number to each of the sample sizes. Those numbers were 5, 10, 20, and 25.

\begin{table}
\begin{tabular}{c c c c c c c c c c}
  \hline
    & data setting               & $n_{11}$ & (SD) & $n_{12}$ & (SD) & $n_{21}$ & (SD) & $n_{22}$ & (SD) \\
  \hline
  1 & balanced and homoscedastic & 5 & (1.0) & 5 & (1.0) & 5  & (1.0) & 5  & (1.0) \\
  2 & differing sample sizes     & 5 & (1.0) & 7 & (1.0) & 10 & (1.0) & 15 & (1.0) \\
  3 & differing variances        & 5 & (1.0) & 5 & (1.3) & 5  & (1.5) & 5  & (2.0) \\
  4 & positive pairings          & 5 & (1.0) & 7 & (1.3) & 10 & (1.5) & 15 & (2.0) \\
  5 & negative pairings          & 5 & (2.0) & 7 & (1.5) & 10 & (1.3) & 15 & (1.0) \\
  \hline
\end{tabular}
\caption{Different subsample sizes and variances considered in the simulation study. Besides the data settings in the table, bigger samples were achieved by adding 5, 10, 20, or 25 observations to each subsample.\label{design}}
\end{table}

There was no effect in the data (i.e. for Equation \ref{eq_data} $\mu=\alpha_i=\beta_j=0$). For the error terms, different symmetric and skewed distributions were used:
\begin{itemize}
  \item Symmetrical distributions: normal, 
  Laplace, 
  logistic,
  and a ``mixed'' distribution,  where each factor level combination has a different symmetric distribution (normal, Laplace, logistic and uniform). 

  \item Skewed distributions: log-normal,
  $\chi_3^2$,
  $\chi_{10}^2$,
   and a ``mixed'' distribution,  where each factor level combination has a different skewed distribution  (exponential, log-normal, $\chi^2_{3}$, $\chi^2_{10}$).
\end{itemize}

To generate variance heterogeneity, random variables were first generated from the distributions mentioned above and standardized to achieve an expected value of 0 and a standard deviation of 1. These values were further multiplied by the standard deviations given in Table \ref{design} to achieve different degrees of variance heteroscedasticity.

\subsubsection{Results}

\begin{figure}
\subfloat{\includegraphics[width=0.49\textwidth,trim=0cm 1.7cm 0cm 0cm,clip=true]{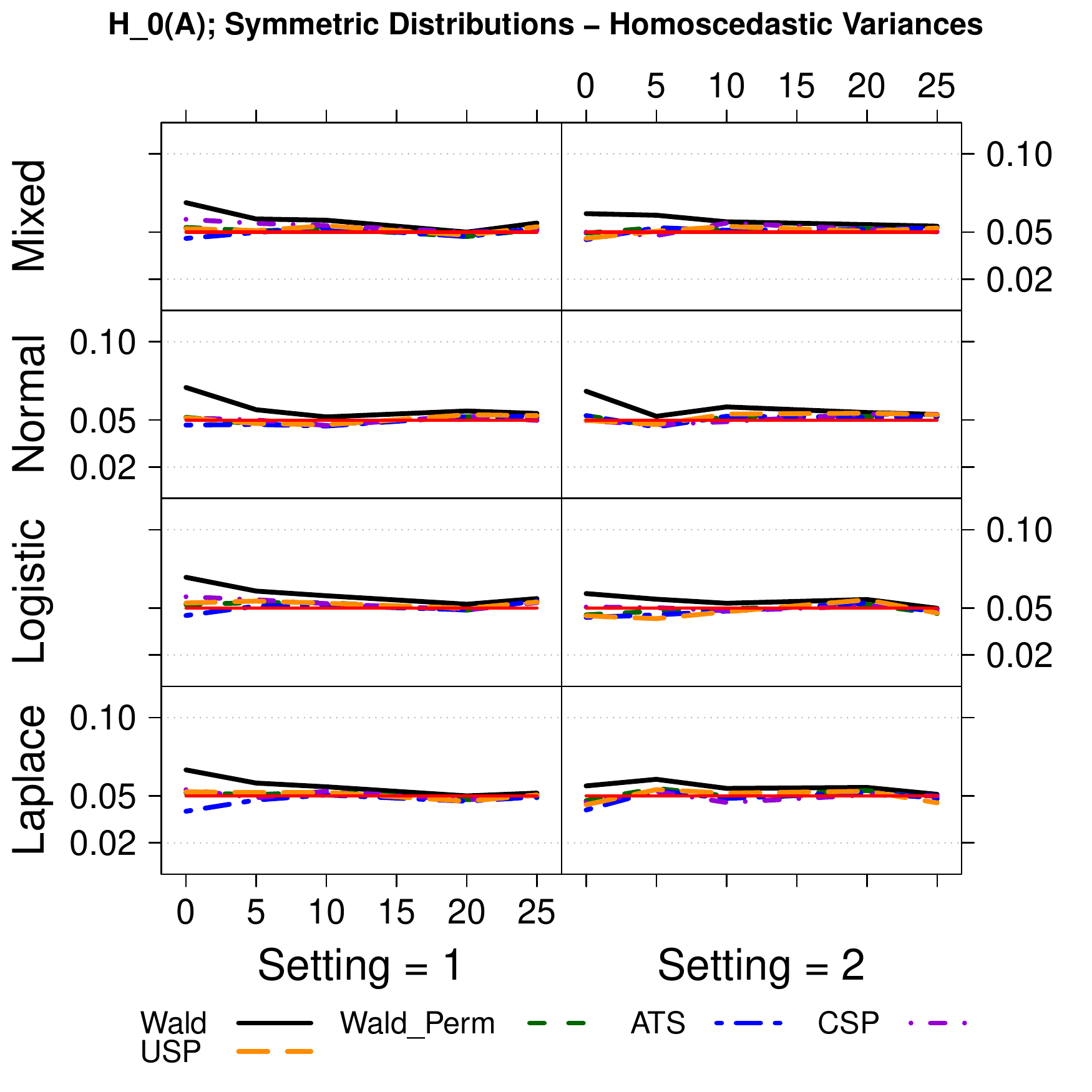}}\hfill
\subfloat{\includegraphics[width=0.49\textwidth,trim=0cm 1.7cm 0cm 0cm,clip=true]{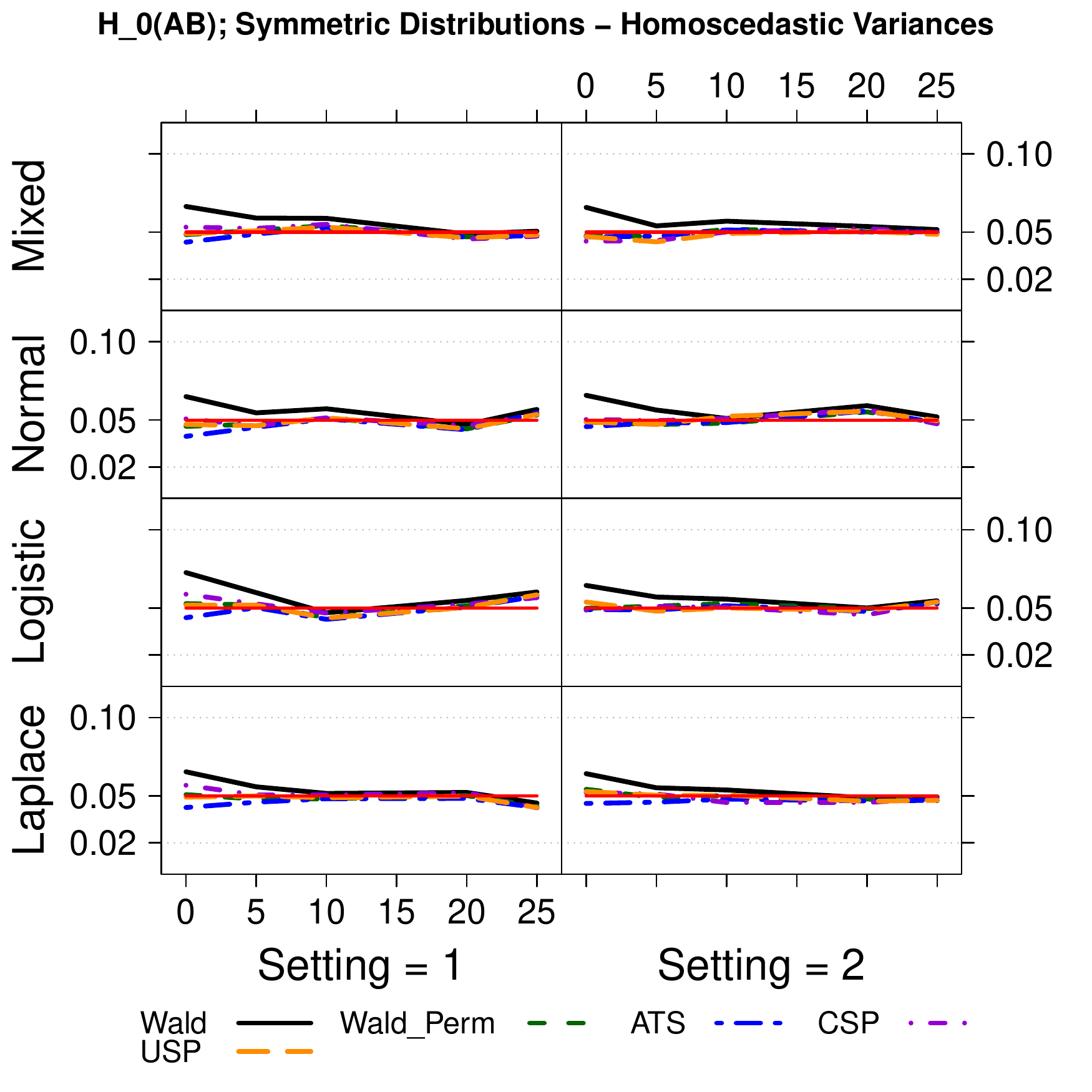}}\\
\subfloat{\includegraphics[width=\textwidth]{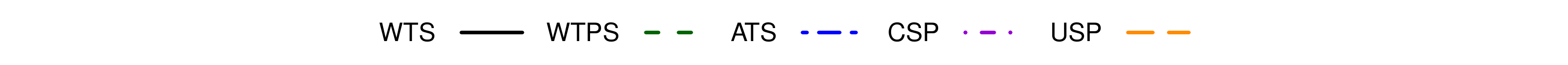}}
\caption{Results for the different procedures testing main effect A (left hand side) or the interaction effect (right hand side) for symmetric distributions and homoscedastic variances. \label{fig:H0SymHom}}
\subfloat{\includegraphics[width=0.49\textwidth,trim=0cm 1.7cm 0cm 0cm,clip=true]{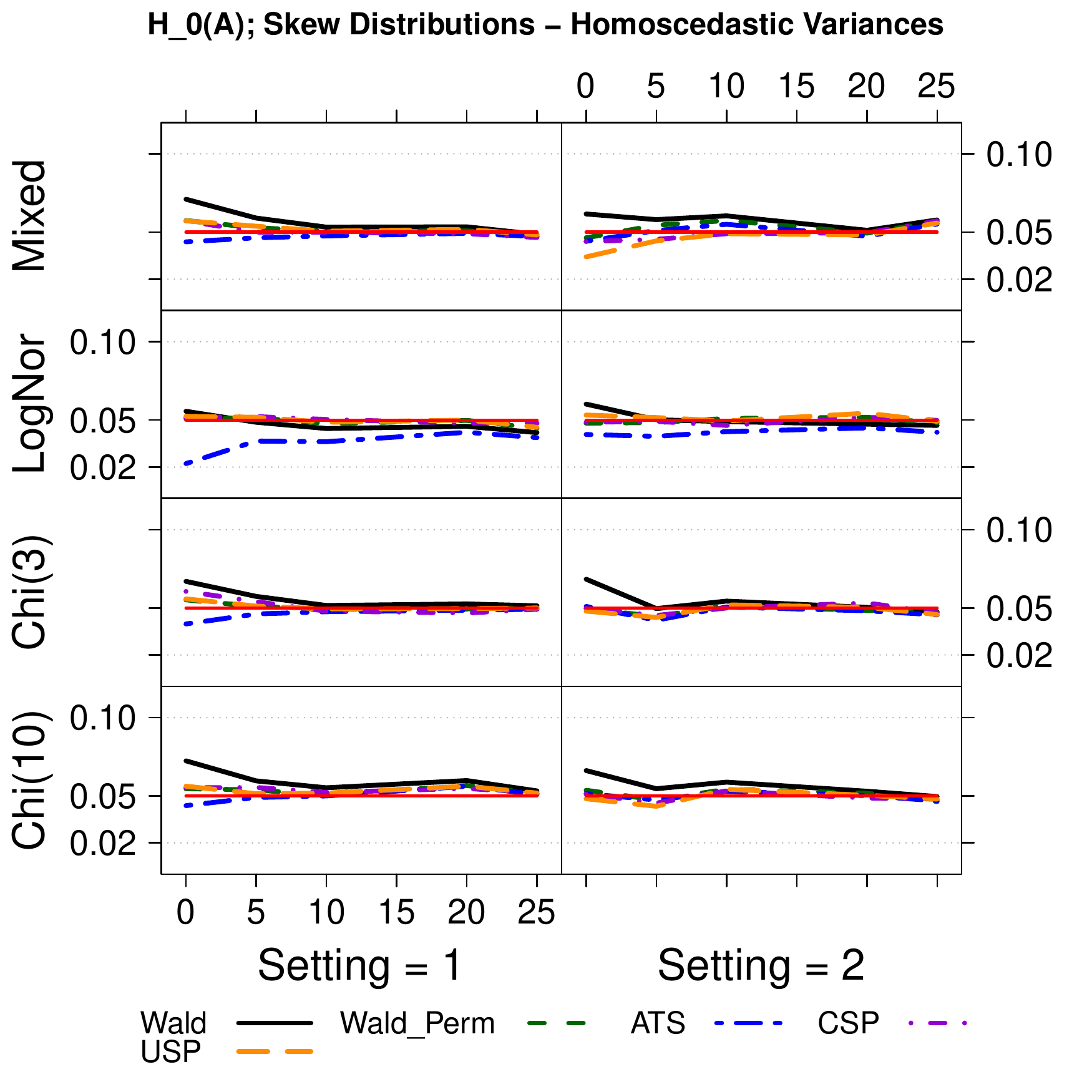}}\hfill
\subfloat{\includegraphics[width=0.49\textwidth,trim=0cm 1.7cm 0cm 0cm,clip=true]{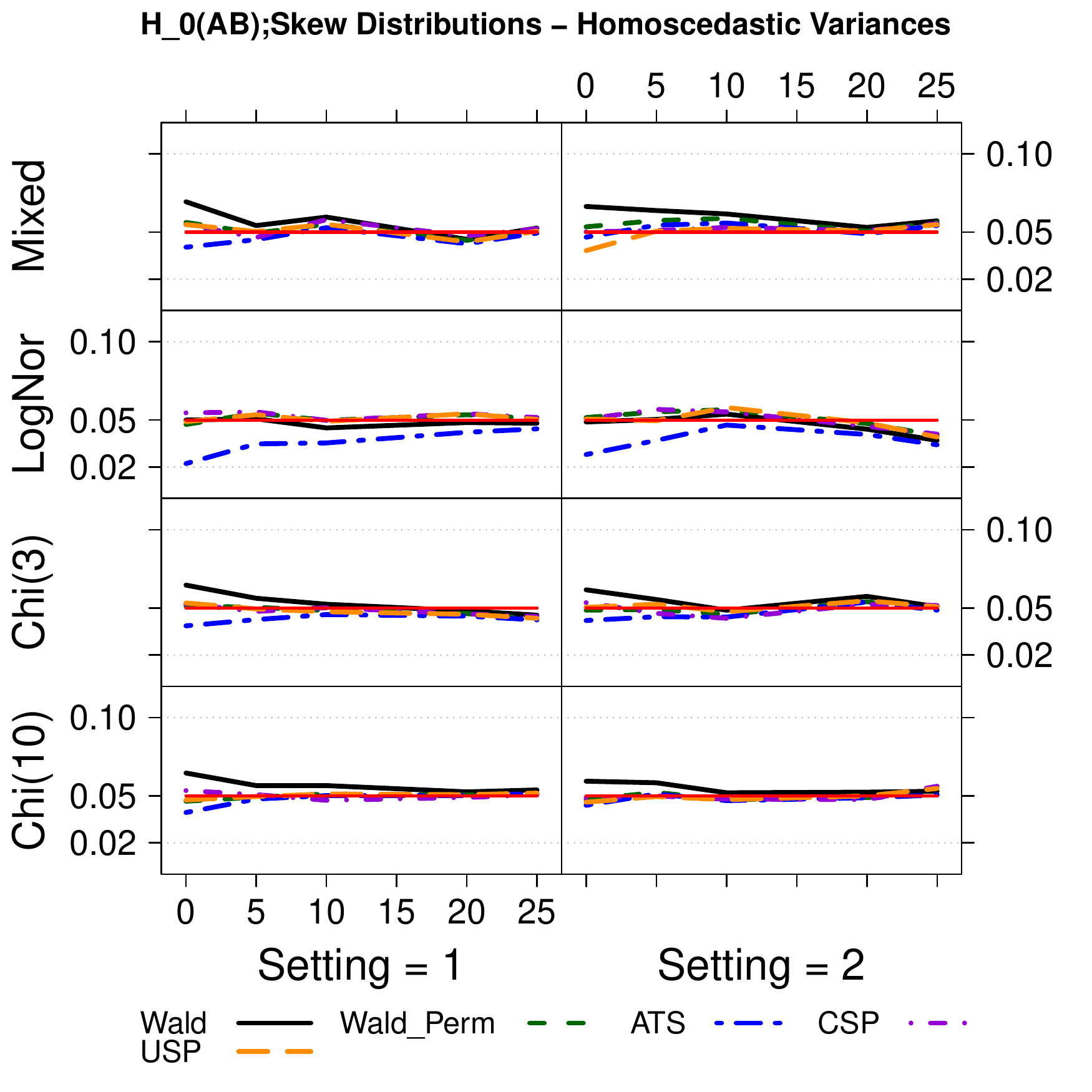}}\\
\subfloat{\includegraphics[width=\textwidth]{legend.pdf}}
\caption{Results for the different procedures testing main effect A (left hand side) or the interaction effect (right hand side) for skewed distributions and homoscedastic variances. \label{fig:H0SkewHom}}
\end{figure}

Figure \ref{fig:H0SymHom} shows the behavior of the different procedures in the case of symmetric and homoscedastic error terms. Most procedures keep close to the nominal $\alpha$-level of .05 that is indicated by the red thin line. WTS tends to be quite liberal, while ATS tends to be slightly conservative for small sample sizes.

Figure \ref{fig:H0SkewHom} shows the behavior for skewed but still homoscedastic error term distributions. The picture is very similar to the previous one, but the conservative behavior of the ATS procedure is more pronounced.

\begin{figure}
\subfloat{\includegraphics[width=0.49\textwidth,trim=0cm 1.7cm 0cm 0cm,clip=true]{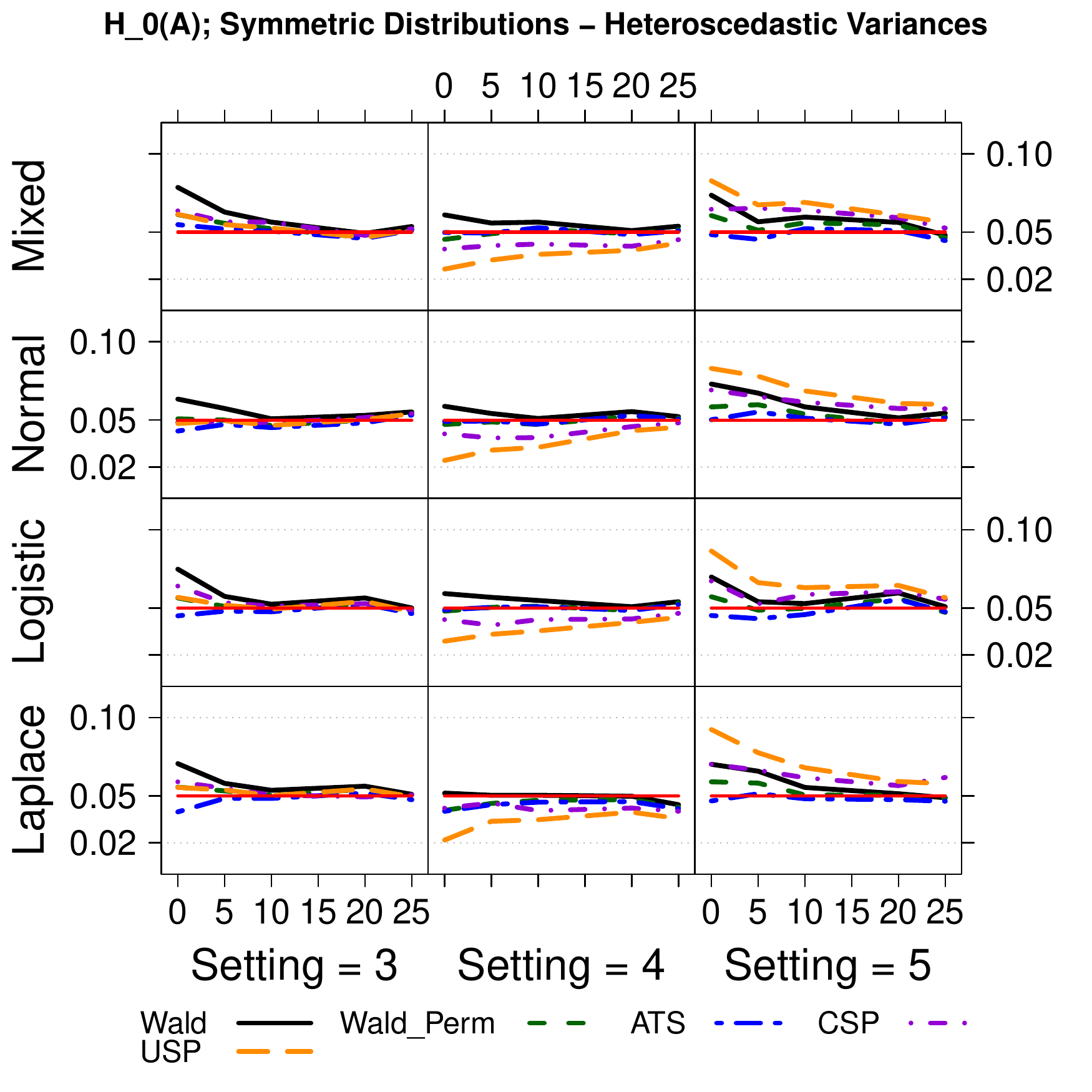}}\hfill
\subfloat{\includegraphics[width=0.49\textwidth,trim=0cm 1.7cm 0cm 0cm,clip=true]{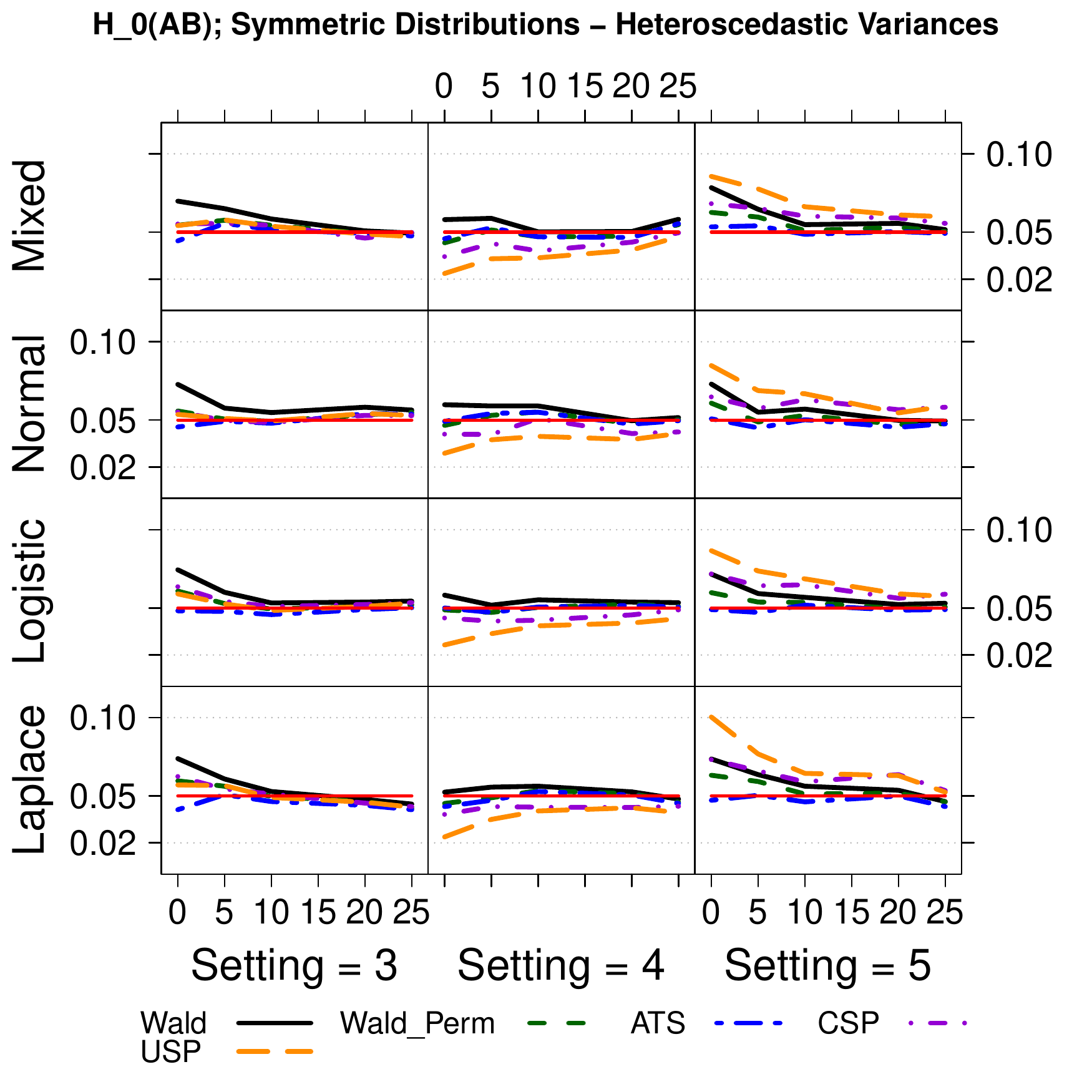}}\\
\subfloat{\includegraphics[width=\textwidth]{legend.pdf}}
\caption{Results for the different procedures testing main effect A (left hand side) or the interaction effect (right hand side) for symmetric distributions and heteroscedastic variances. \label{fig:H0SymHet}}
\subfloat{\includegraphics[width=0.49\textwidth,trim=0cm 1.7cm 0cm 0cm,clip=true]{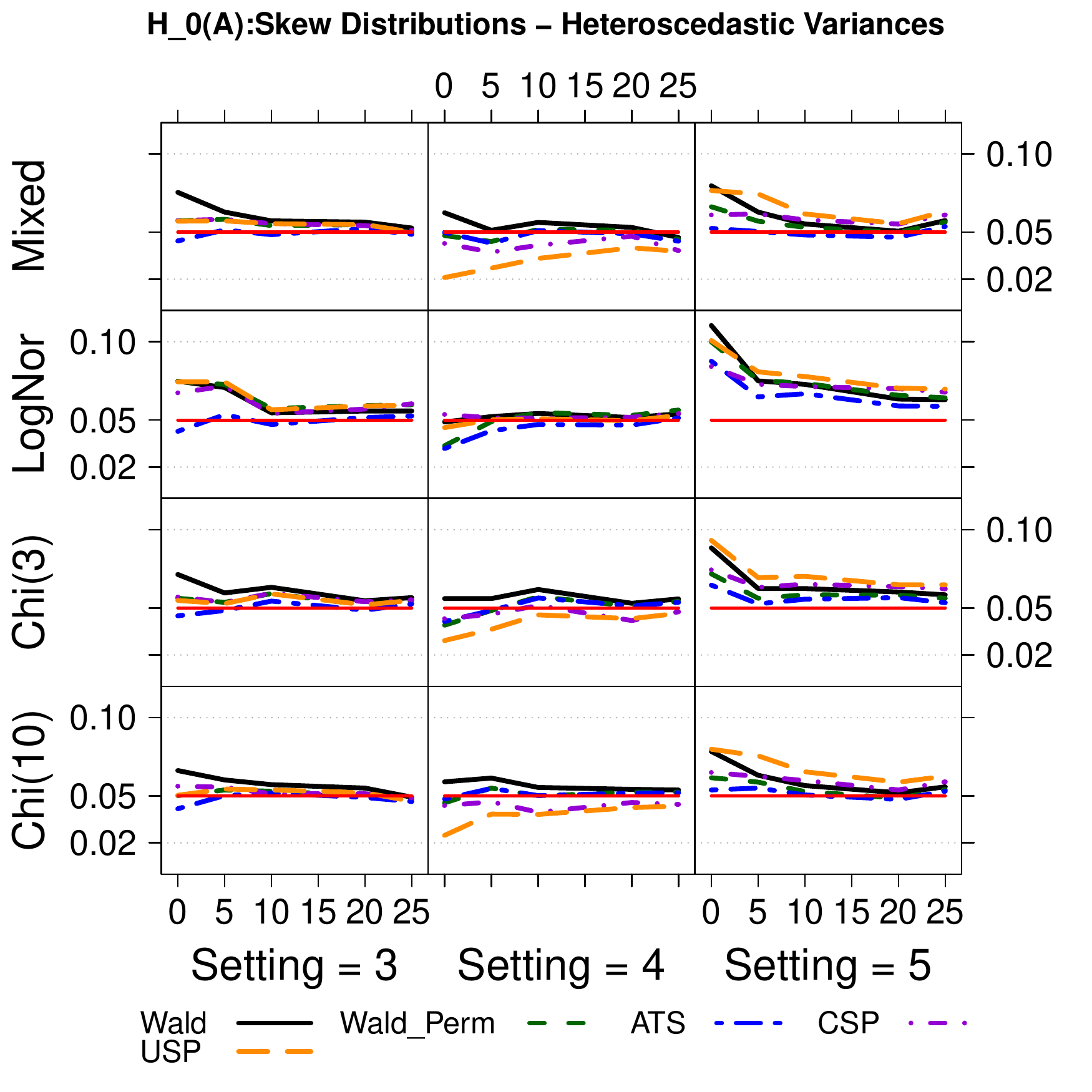}}\hfill
\subfloat{\includegraphics[width=0.49\textwidth,trim=0cm 1.7cm 0cm 0cm,clip=true]{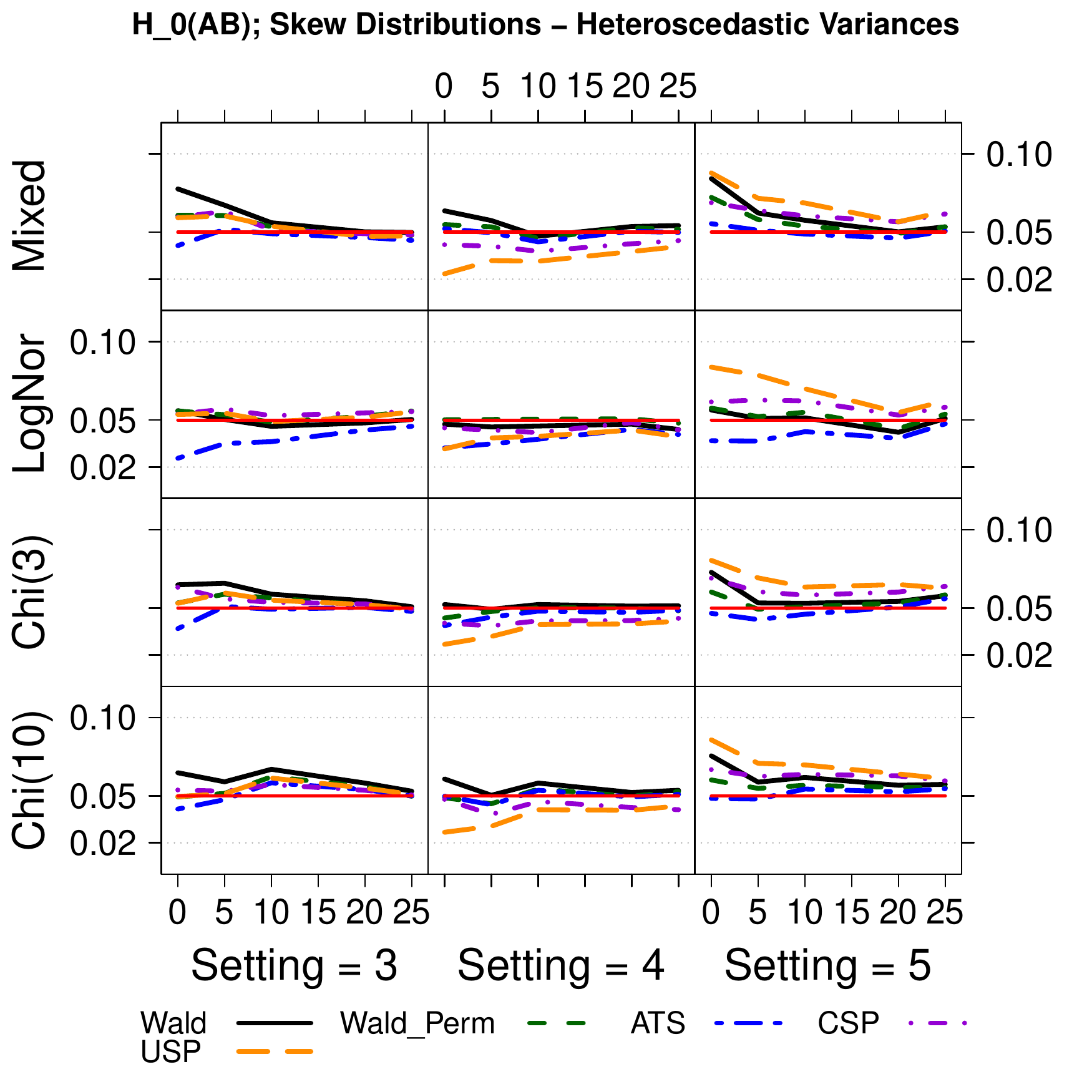}}\\
\subfloat{\includegraphics[width=\textwidth]{legend.pdf}}
\caption{Results for the different procedures testing main effect A (left hand side) or the interaction effect (right hand side) for skewed distributions and heteroscedastic variances. \label{fig:H0SkewHet}}
\end{figure}

Figure \ref{fig:H0SymHet} shows the behavior in the symmetric and heteroscedastic case. For Setting 3 with equal sample sizes there is not much difference in comparison to the previous cases. In Setting 4, the positive pairings, WTPS and ATS show a good adherence to the level and a slightly conservative behavior in the case of the Laplace-distribution. WTS tends to over-reject the null in small sample size settings. Both the CSP and USP tests tend to result in conservative decisions. This is more pronounced for small sample sizes and for the USP-procedure. In Setting 5, that indicates negative pairings, all procedures unless ATS tend to result in a liberal behavior--especially for small sample sizes. USP has the strongest tendency with Type-I-error rates up to .08.

Figure \ref{fig:H0SkewHet} shows the behavior in the skewed and heteroscedastic case. In general, the same conclusions can be drawn. For the log-normal distribution there is a general tendency to get a more liberal decision than in the other cases. This means that in Setting 4 with positive pairings the procedures keep the level almost well, but in the other cases the Type-I-error rate is up to .10.

\subsection{Data Sets Containing an Effect}

\subsubsection{Description}
Table \ref{design_power} shows the different combinations of subsample sizes and standard deviations for data sets that contained an effect. Two aspects were considered: The power behavior as well as the level of the procedures when testing an inactive effect. To ensure a valid comparison of the power behavior, the sample sizes and variance heterogeneity was chosen less extreme than in the previous simulation study (see Section \ref{subsec:NoEffect}).

\begin{table}[b]

\begin{tabular}{c c c c c c c c c c}
  \hline
  &data setting & $n_{11}$ & (SD) & $n_{12}$ & (SD) & $n_{21}$ & (SD) & $n_{22}$ & (SD) \\
  \hline
  1 & balanced and homoscedastic & 10 & (1)             & 10 & (1)             & 10 & (1)             & 10 & (1) \\
  2 & differing sample sizes     &  9 & (1)             &  9 & (1)             & 15 & (1)             & 15 & (1) \\
  3 & differing variances        & 10 & (1)             & 10 & (1)             & 10 & ($\sqrt[4]{2}$) & 10 & ($\sqrt[4]{2}$) \\
  4 & positive pairings          &  9 & (1)             &  9 & (1)             & 15 & ($\sqrt[4]{2}$) & 15 & ($\sqrt[4]{2}$) \\
  5 & negative pairings          &  9 & ($\sqrt[4]{2}$) &  9 & ($\sqrt[4]{2}$) & 15 & (1)             & 15 & (1) \\
  \hline
\end{tabular}
\caption{Different subsample sizes and standard deviations considered in the simulation study containing effects. \label{design_power}}
\end{table}

Again different error term distributions were used:
\begin{itemize}
  \item normal and Laplace distribution as symmetric distributions, and
  \item log-normal distribution and exponential distribution as skewed distributions.
\end{itemize}

Different error term variances were obtained in the same manner as described above in Section \ref{subsec:NoEffect} using the standard deviations from Table \ref{design_power}.
Additionally, there were active effects as described in Table \ref{effects_power} in the data with $\mu=0$ and $\delta\in\{0,0.2,\dots,1\}$. The tested effects were again main effect A and the interaction effect. In some cases where only main effect B was active the aim was to test if the procedures kept the level in these cases.
\begin{table}
\begin{tabular}{c c c c c c c c c p{4cm}}
  \hline
  Condition & $\alpha_1$ & $\alpha_2$ & $\beta_1$ & $\beta_2$ & $\alpha\beta_{11}$ & $\alpha\beta_{12}$ & $\alpha\beta_{21}$ & $\alpha\beta_{22}$ & active effects \\
  \hline
  1         & $+\delta$  & $-\delta$  & 0         & 0         & 0                  & 0                  & 0                  & 0 &main effect A\\
  2         & 0          & 0          & $+\delta$ & $-\delta$ & 0                  & 0                  & 0                  & 0 &main effect B\\
  3         & $+\frac{\delta}{2}$ & $-\frac{\delta}{2}$ & 0 & 0 & $+\frac{\delta}{2}$ & $-\frac{\delta}{2}$ & $-\frac{\delta}{2}$ & $+\frac{\delta}{2}$ & main effect A and interaction effect \\
  \hline
\end{tabular}
\caption{Different effect in simulated data sets with $\mu=0$ and $\delta\in\{0,0.2,\dots,1\}$ in the simulation study containing effects. \label{effects_power}}
\end{table}

\subsubsection{Results}

\begin{figure}
\subfloat{\includegraphics[width=0.33\textwidth,trim=0cm 1.7cm 0cm 0cm,clip=true]{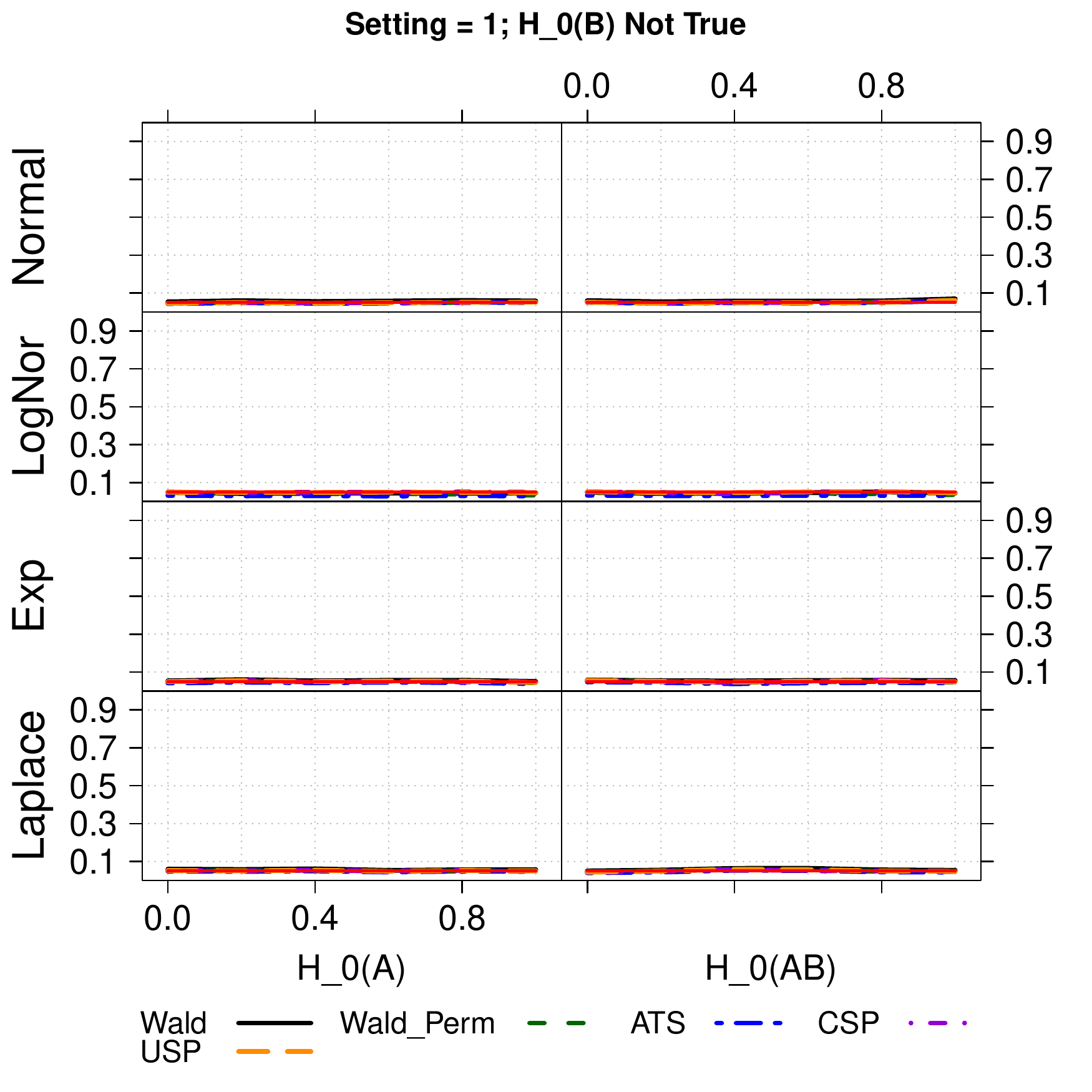}}\hfill
\subfloat{\includegraphics[width=0.33\textwidth,trim=0cm 1.7cm 0cm 0cm,clip=true]{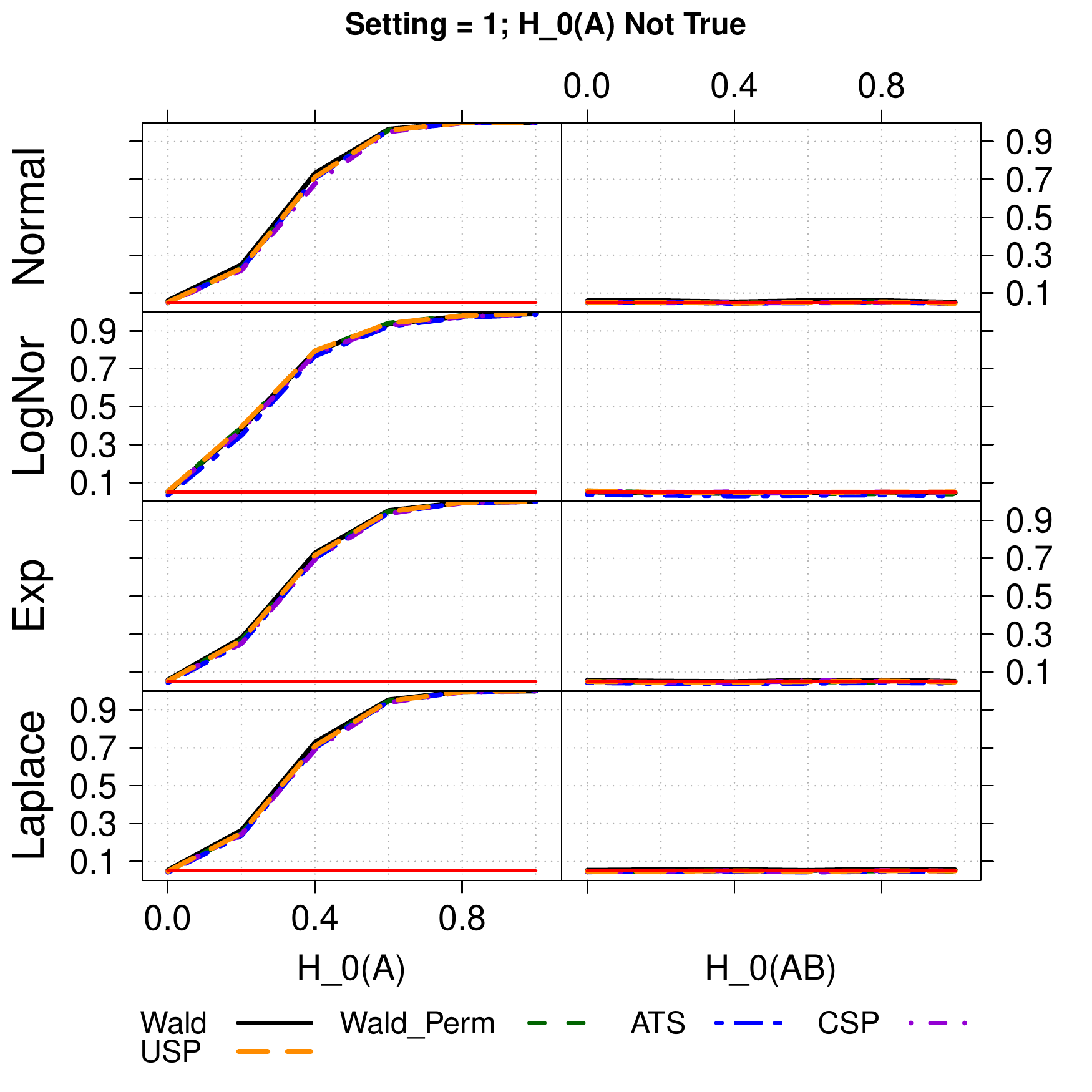}}\hfill
\subfloat{\includegraphics[width=0.33\textwidth,trim=0cm 1.7cm 0cm 0cm,clip=true]{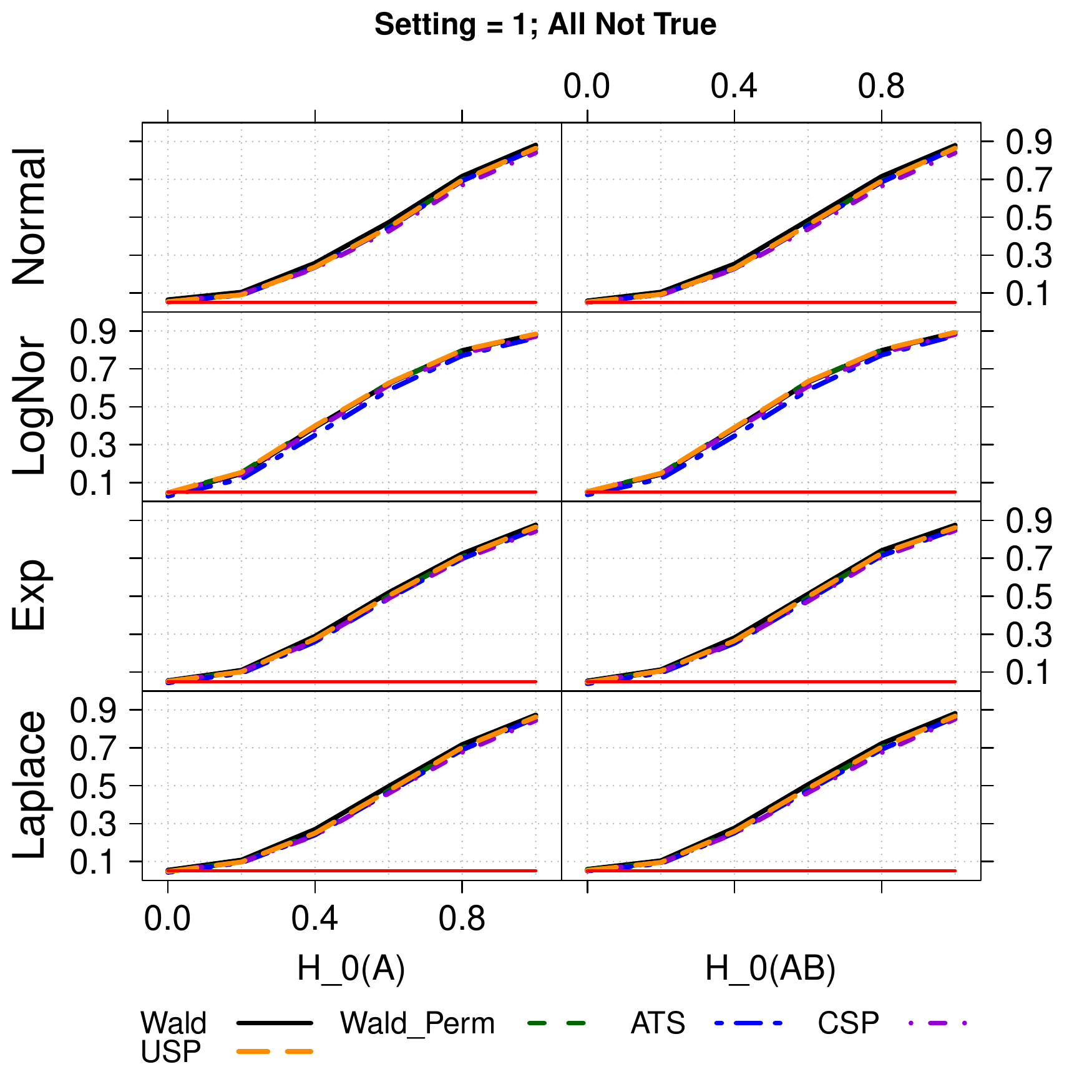}}\\
\subfloat{\includegraphics[width=\textwidth]{legend.pdf}}
\caption{Results for data sets containing effects (equal subsample sizes and homoscedastic variances). \label{fig:H1Setting1}}

\subfloat{\includegraphics[width=0.33\textwidth,trim=0cm 1.7cm 0cm 0cm,clip=true]{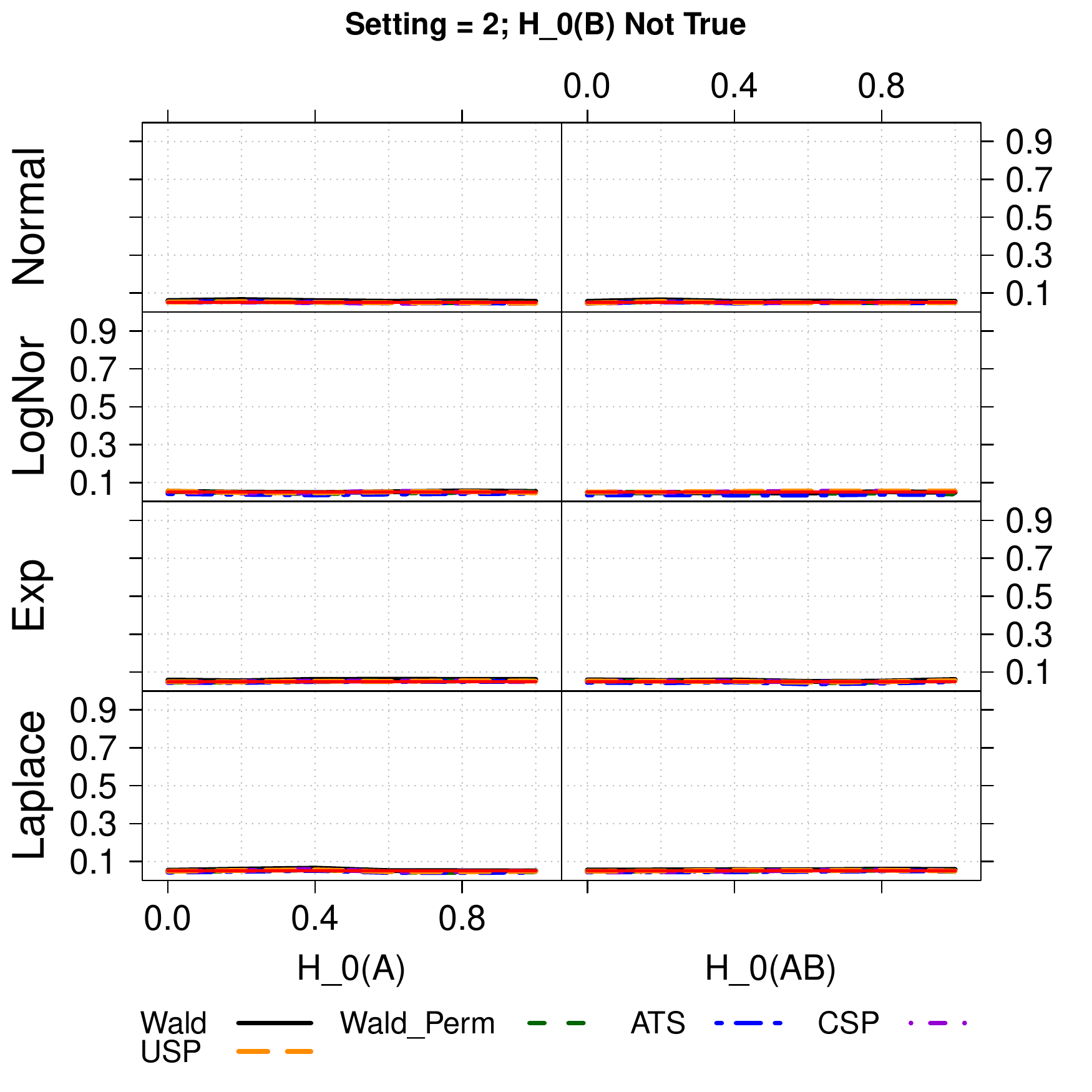}}\hfill
\subfloat{\includegraphics[width=0.33\textwidth,trim=0cm 1.7cm 0cm 0cm,clip=true]{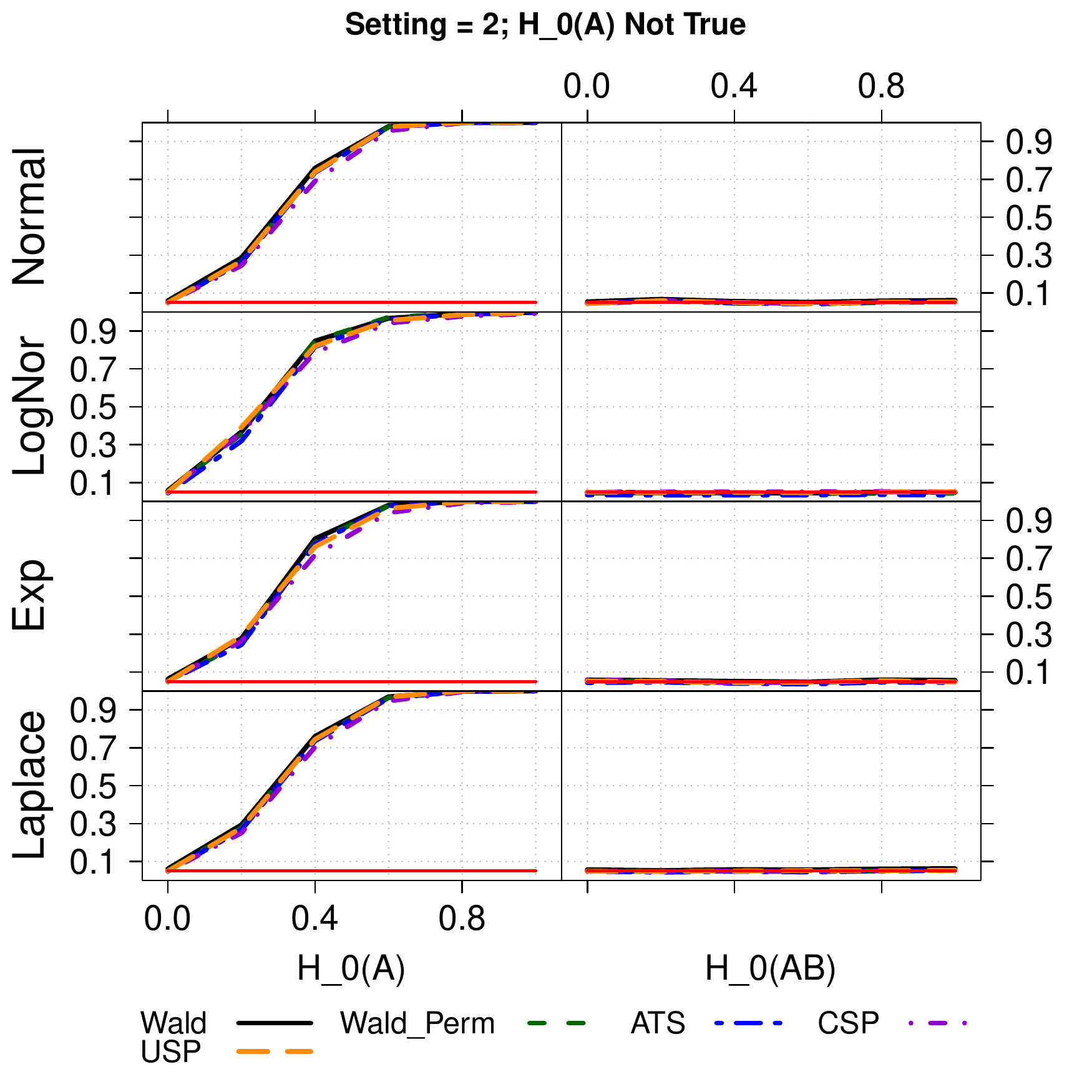}}\hfill
\subfloat{\includegraphics[width=0.33\textwidth,trim=0cm 1.7cm 0cm 0cm,clip=true]{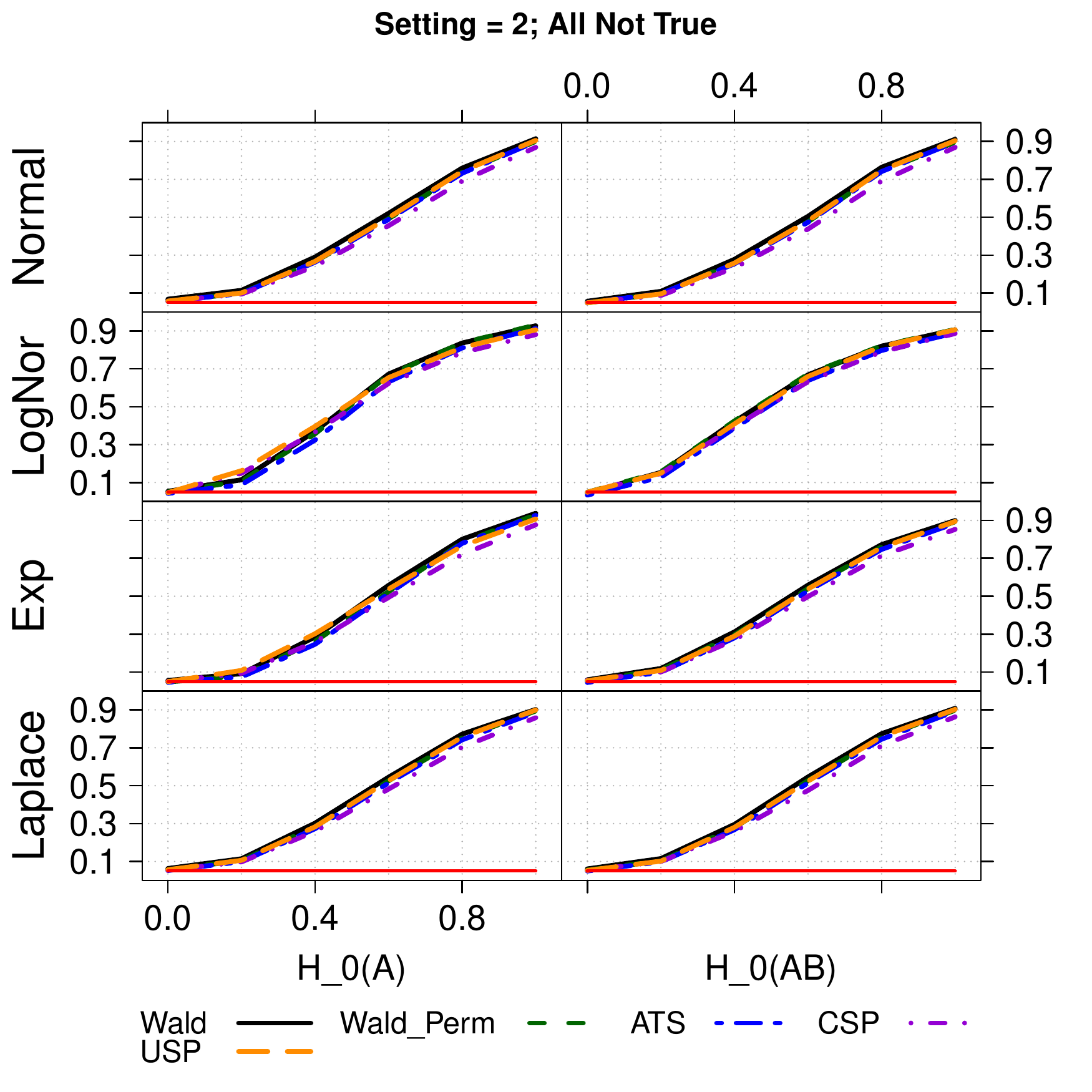}}\\
\subfloat{\includegraphics[width=\textwidth]{legend.pdf}}
\caption{Results for data sets containing effects (equal subsample sizes and heteroscedastic variances). \label{fig:H1Setting2}}
\end{figure}

Figure \ref{fig:H1Setting1}--\ref{fig:H1Setting1} show the behavior of the different procedures for data sets containing an effect. The procedures show a very similar power behavior.

\begin{figure}
\subfloat{\includegraphics[width=0.33\textwidth,trim=0cm 1.7cm 0cm 0cm,clip=true]{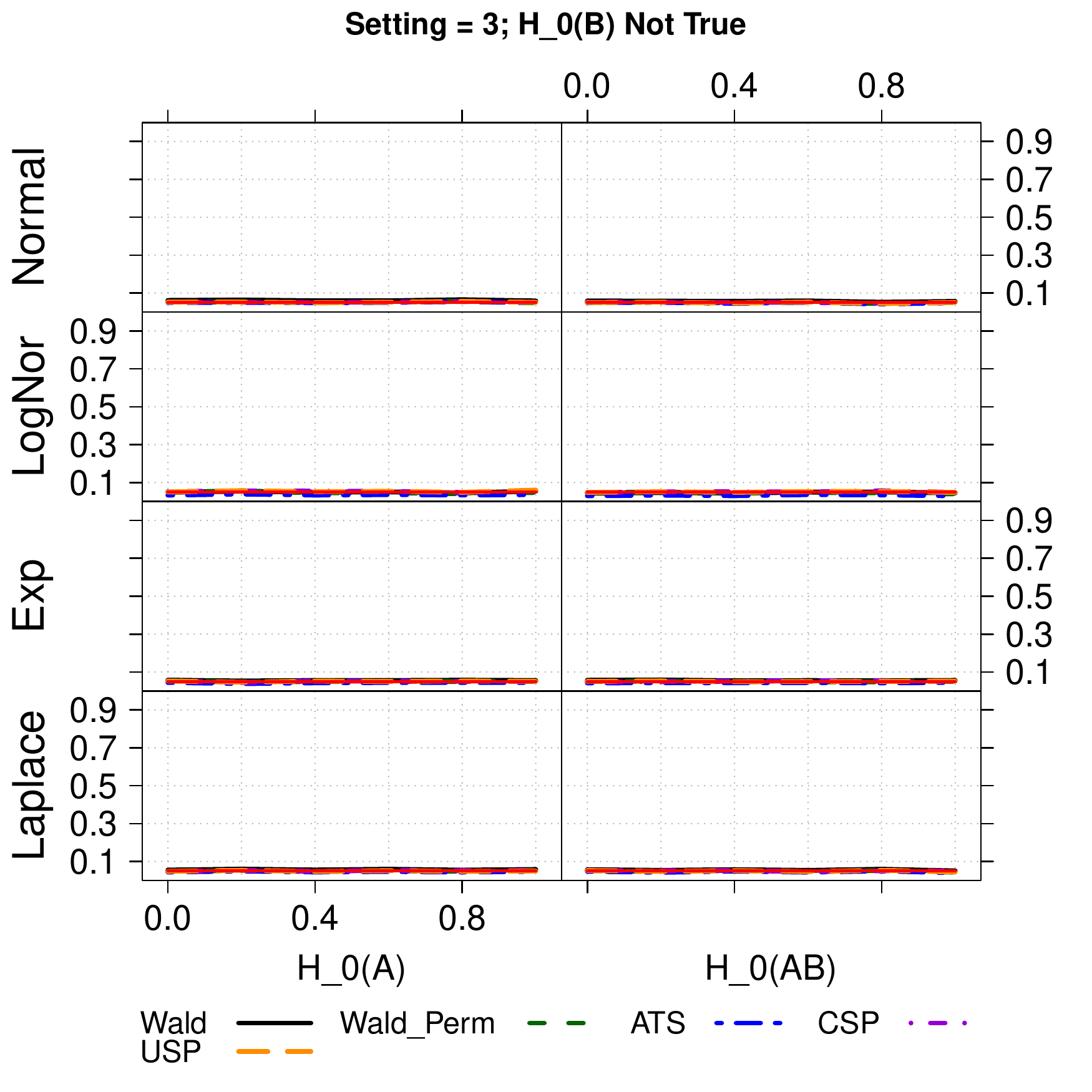}}\hfill
\subfloat{\includegraphics[width=0.33\textwidth,trim=0cm 1.7cm 0cm 0cm,clip=true]{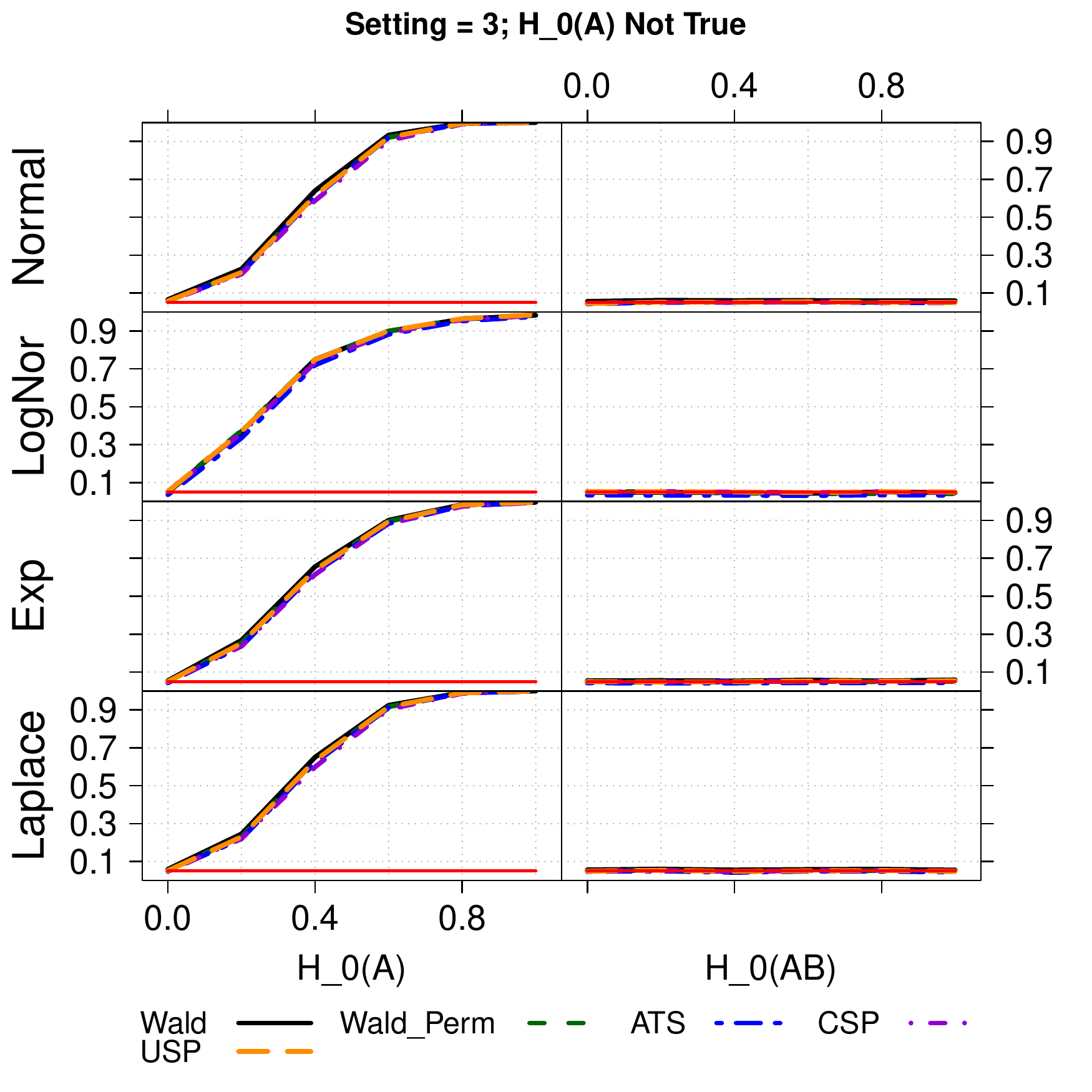}}\hfill
\subfloat{\includegraphics[width=0.33\textwidth,trim=0cm 1.7cm 0cm 0cm,clip=true]{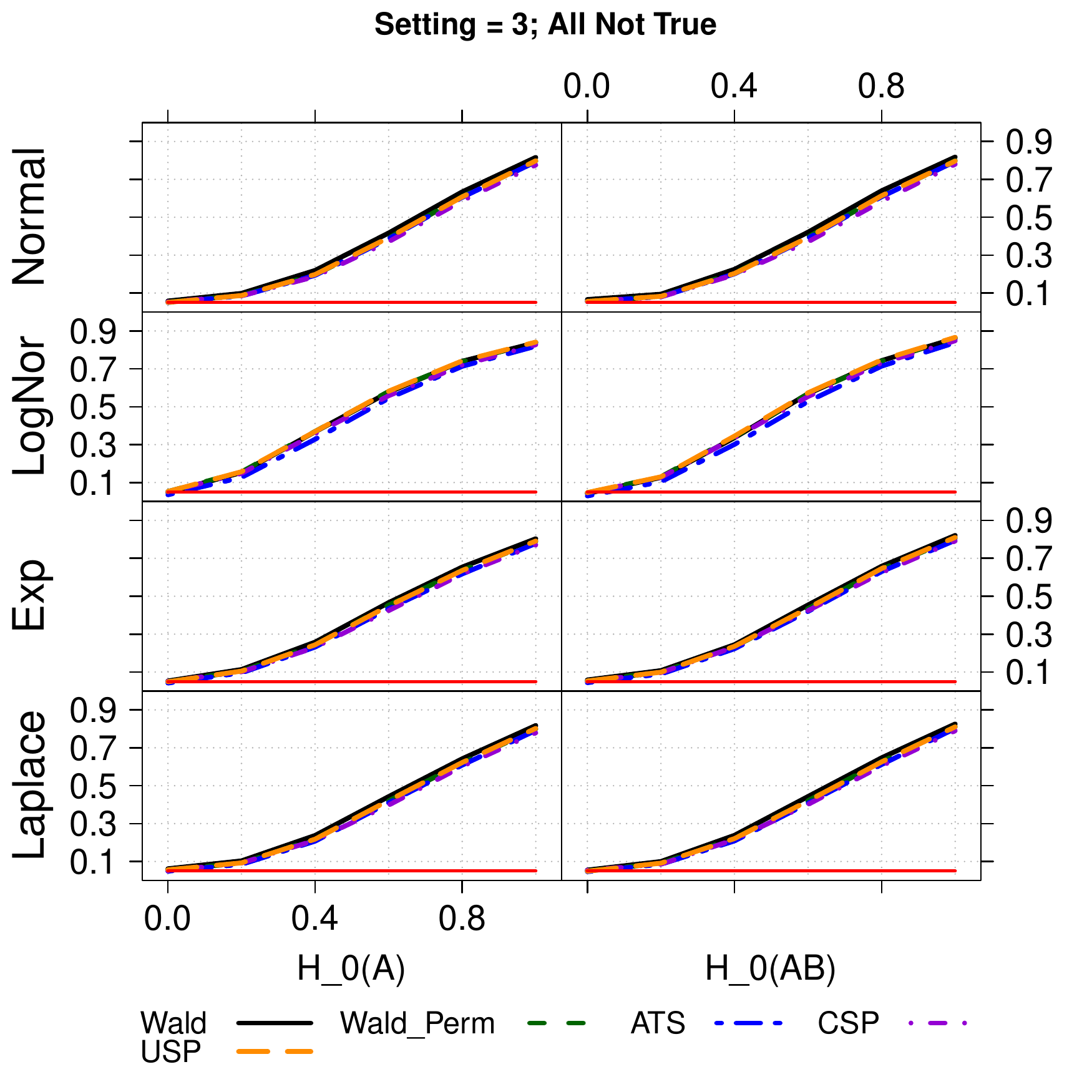}}\\
\subfloat{\includegraphics[width=\textwidth]{legend.pdf}}
\caption{Results for data sets containing effects (unequal subsample sizes and homoscedastic variances). \label{fig:H1Setting3}}

\subfloat{\includegraphics[width=0.33\textwidth,trim=0cm 1.7cm 0cm 0cm,clip=true]{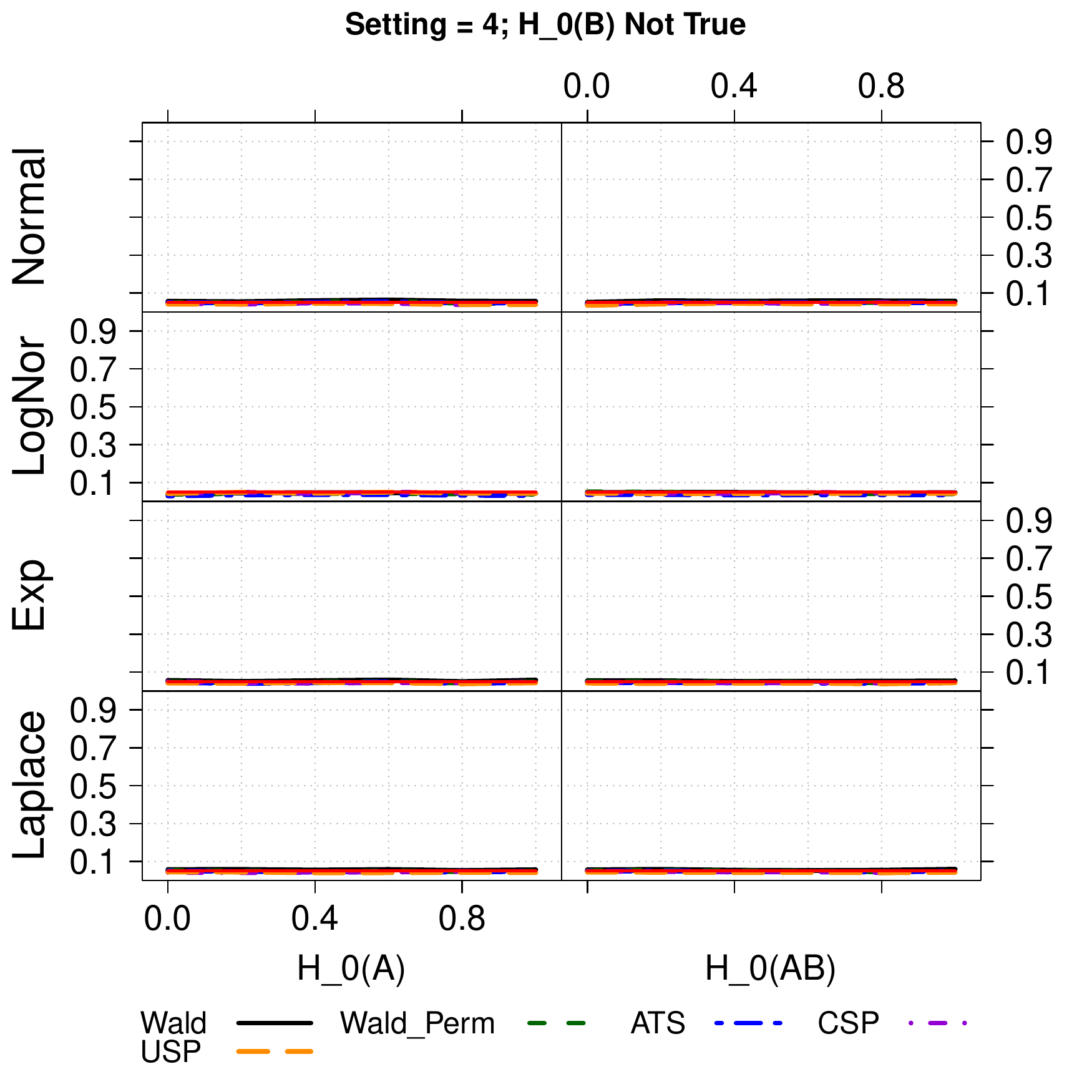}}\hfill
\subfloat{\includegraphics[width=0.33\textwidth,trim=0cm 1.7cm 0cm 0cm,clip=true]{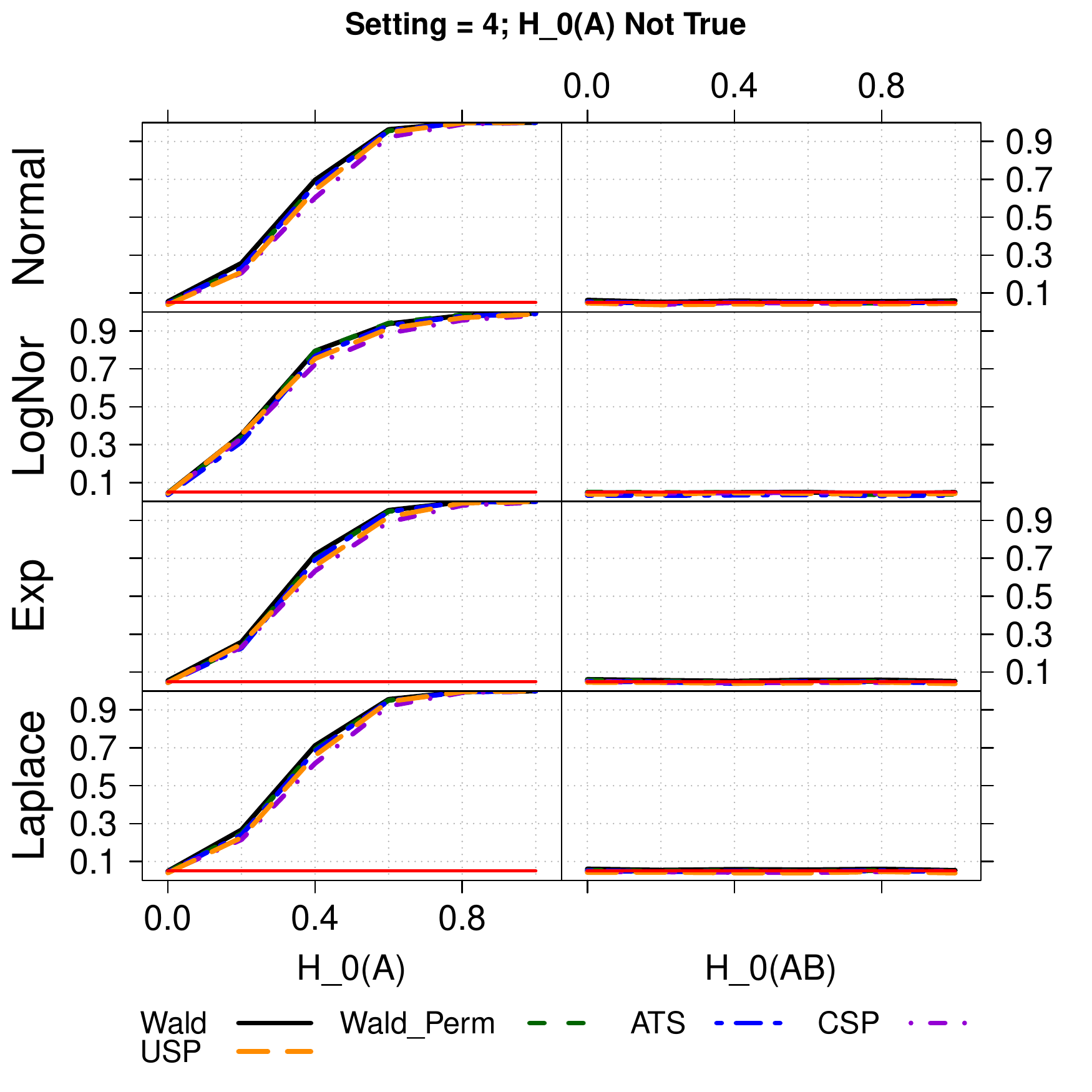}}\hfill
\subfloat{\includegraphics[width=0.33\textwidth,trim=0cm 1.7cm 0cm 0cm,clip=true]{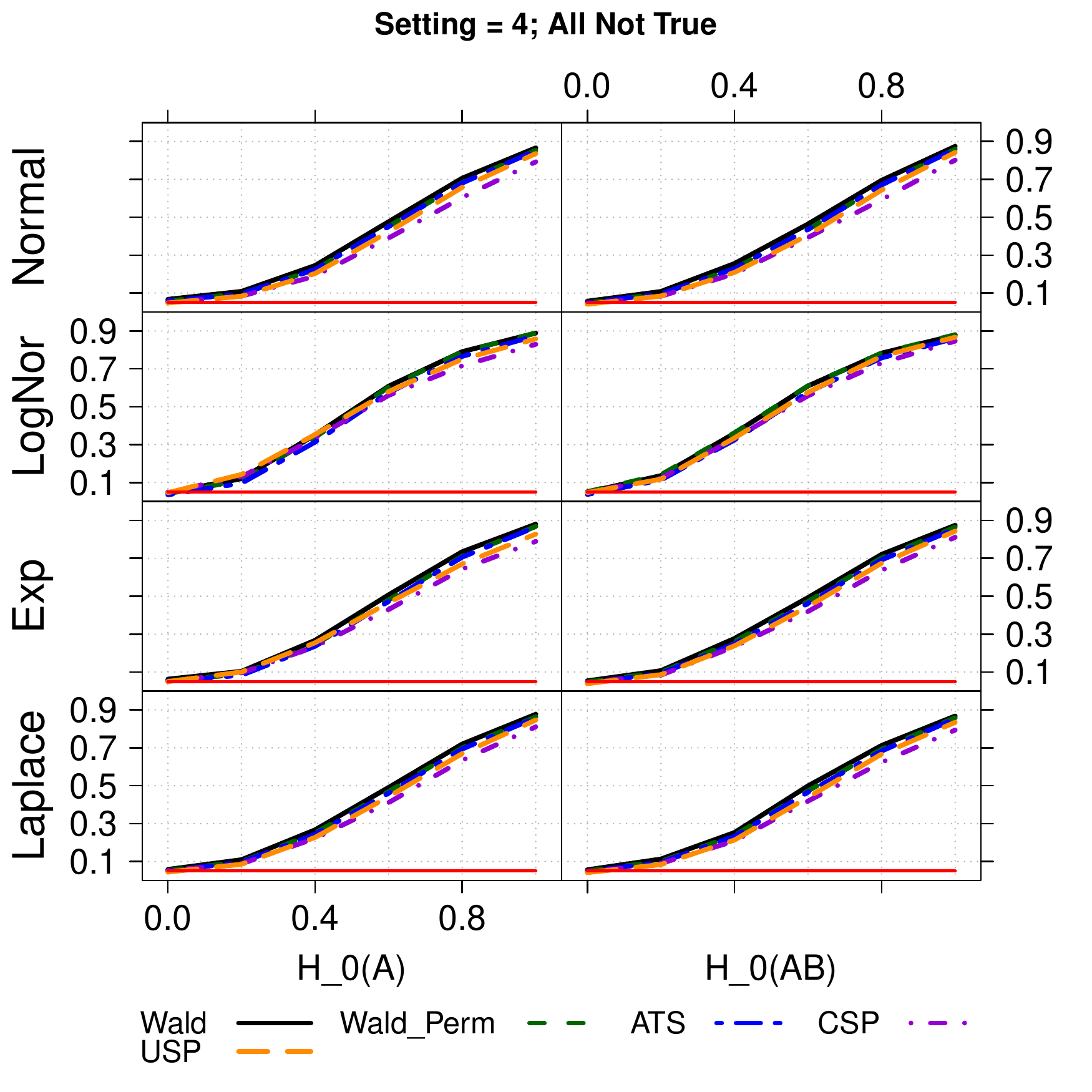}}\\
\subfloat{\includegraphics[width=\textwidth]{legend.pdf}}
\caption{Results for data sets containing effects (positive pairings). \label{fig:H1Setting4}}

\subfloat{\includegraphics[width=0.31\textwidth,trim=0cm 1.7cm 0cm 0cm,clip=true]{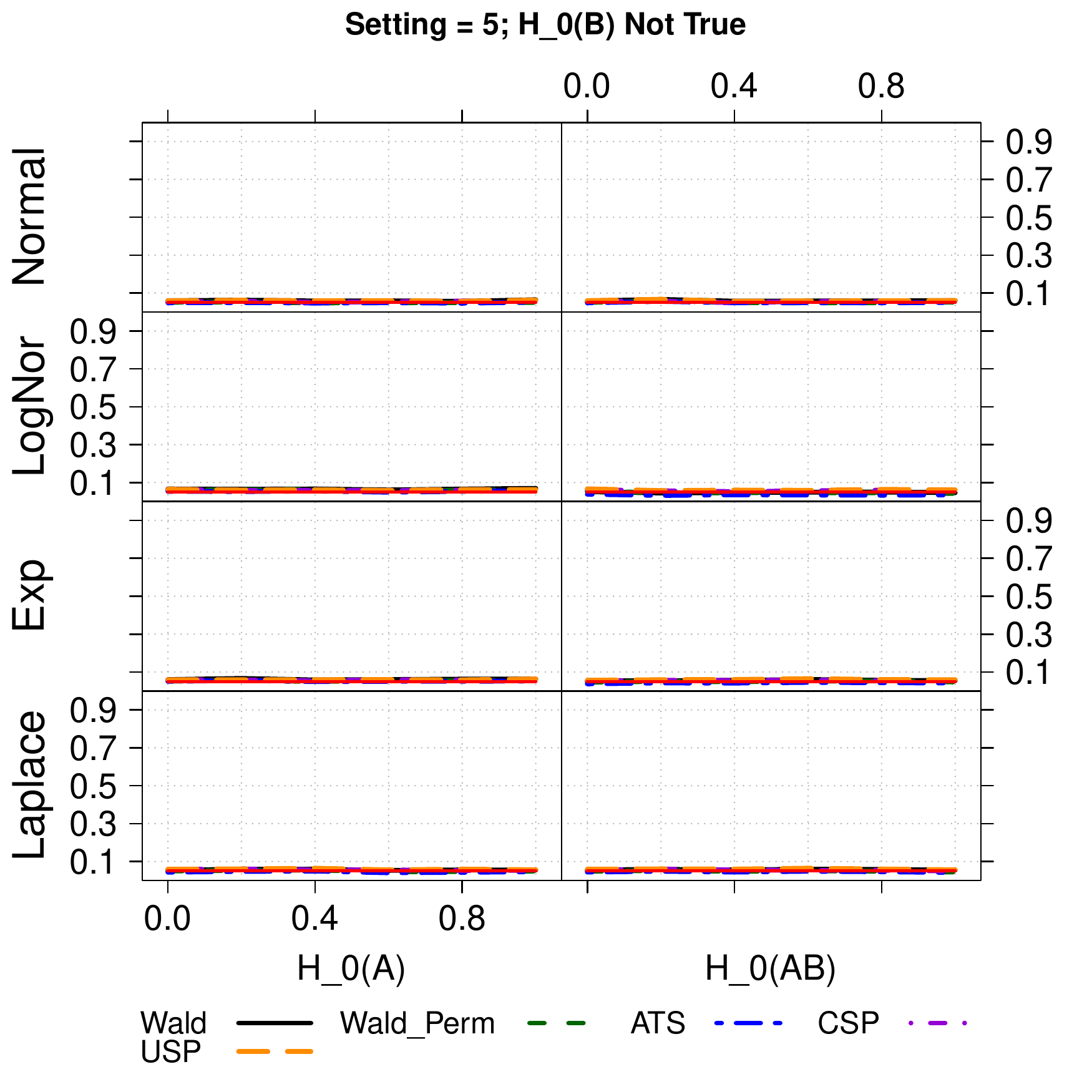}}\hfill
\subfloat{\includegraphics[width=0.31\textwidth,trim=0cm 1.7cm 0cm 0cm,clip=true]{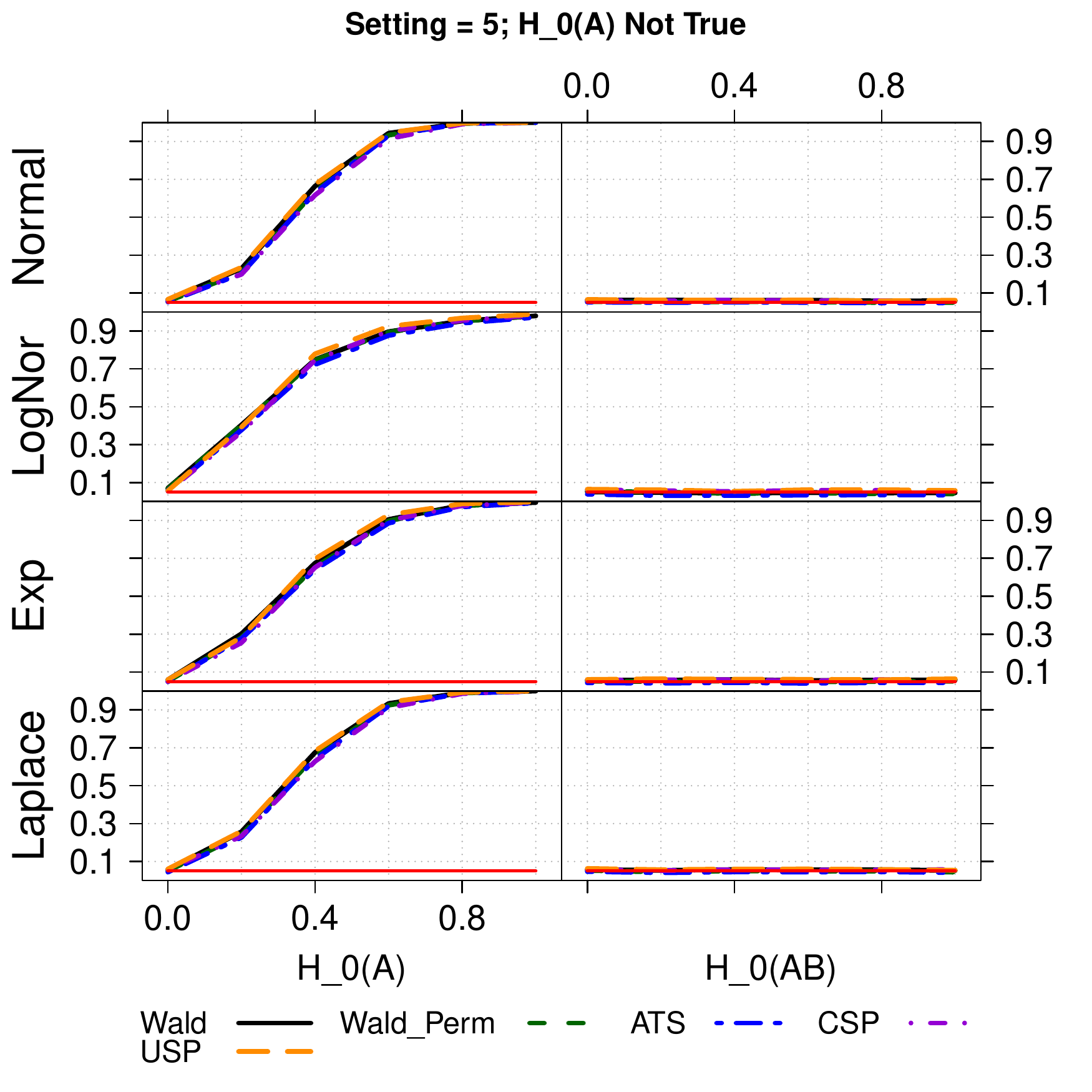}}\hfill
\subfloat{\includegraphics[width=0.31\textwidth,trim=0cm 1.7cm 0cm 0cm,clip=true]{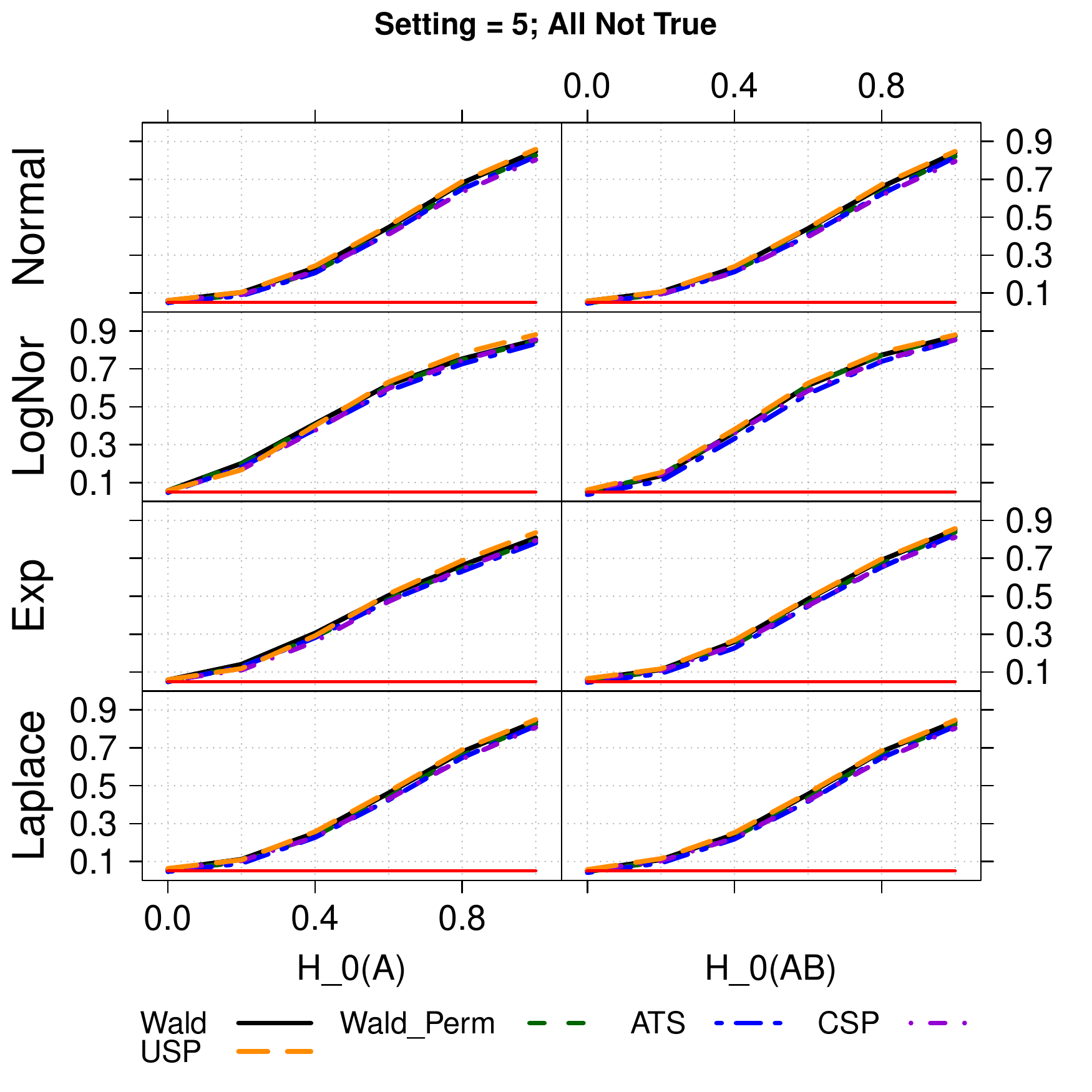}}\\
\subfloat{\includegraphics[width=\textwidth]{legend.pdf}}
\caption{Results for data sets containing effects (negative pairings). \label{fig:H1Setting5}}
\end{figure}

\section{Conclusion}
\label{sec:conclusion}

As the simulation study showed, the different procedures may be useful depending on the data setting and further aspects.

The ATS procedure was the only one that never exceeded the nominal level. On the other hand it may show a conservative behavior, but in the simulations containing effects this was only slightly observable. Similar to the results of previous simulation studies, the conservative behavior was higher for skewed distributions, especially with homoscedastic error term variances. An advantage of this procedure is that it can be adapted for very different designs and hypotheses \citep[see][for more background information]{di253}.

The WTS procedure showed in almost every data setting a liberal behavior for small samples. It should be only applied when sample sizes are large.

The WTPS procedure overcomes this problem. In all considered simulation settings this procedure controls the type-I error rate quite accurately. In case of positive or negative pairings, this permutation test shows better results than its competitors.
Both the WTS and WTPS can be adapted to higher-way layouts and hierarchical designs. \\
The CSP and the USP procedures work well for all cases with equal subsample sizes or homogeneous variances. This implies cases where exchangeability of the observations might not be given due to different error term distributions (mixed distributions) or heterogeneous variances. In case of positive and negative pairings, the behavior is similar to parametric ANOVA with a conservative behavior for positive pairings and a liberal behavior for negative pairings. This is more pronounced for the USP-procedure. The power behavior of both procedures was very comparable to the other procedures. CSP showed in some cases a slightly lower power than the other procedures.
The CSP and the USP procedures are restricted to certain hypothesis due to their construction: They assumed that all cells should get the same weight in the analysis. This corresponds to Type III sums of squares \citep{di154}. Extension of these procedures to other kind of hypotheses, unbalancedness, or more complex designs might be challenging.

There are various aspects at which this line of research might be continued:
\begin{itemize}
  \item The simulation study could be extended by looking at other data settings (e.g., null pairings) or by including further procedures \citep[e.g.,][]{di138}.
  \item The CSP and the USP algorithm are still not widely applicable. So more research on these procedures or possibly combination of these procedures with other approaches could enlarge their scope of application.
	\item We restricted our analysis to two by two ANOVA models. Permutation tests allowing for covariates in higher way layouts will be part of future research.
\end{itemize}

\bibliographystyle{asa}
\bibliography{paper}

\end{document}